\documentclass[12pt]{article}
\pdfoutput=1
\usepackage{dsfont}
\usepackage[DIV13]{typearea}
\usepackage{amsmath}
\usepackage{amssymb}
\usepackage{graphicx}
\usepackage{cite}
\usepackage{bbold}
\newcommand{\be}{\beta}

\def\be{\begin{equation}}
\def\ee{\end{equation}}
\def\beq{\begin{equation}}
\def\eeq{\end{equation}}
\def\bc{\begin{center}}
\def\ec{\end{center}}
\def\bea{\begin{eqnarray}}
\def\eea{\end{eqnarray}}
\def\dd{\displaystyle}
\def\nn{\nonumber}
\begin{document} 
\begin{titlepage}
\vspace*{-1cm}
\phantom{hep-ph/***} 
\flushright
\hfil{DFPD-2017/TH/09}\\
\vskip 1.5cm
\begin{center}
\mathversion{bold}
{\LARGE\bf Are neutrino masses modular forms?}\\[3mm]
\mathversion{normal}
\vskip .3cm
\end{center}
\vskip 0.5  cm
\begin{center}
{\large Ferruccio Feruglio}
\\
\vskip .7cm
{\footnotesize
Dipartimento di Fisica e Astronomia `G.~Galilei', Universit\`a di Padova
\\
INFN, Sezione di Padova, Via Marzolo~8, I-35131 Padua, Italy
\\
\vskip .1cm
\vskip .5cm
\begin{minipage}[l]{.9\textwidth}
\begin{center} 
\textit{E-mail: feruglio@pd.infn.it} 
\end{center}
\end{minipage}
}
\end{center}
\vskip 1cm
\begin{abstract}
We explore a new class of supersymmetric models for lepton masses and mixing angles where the role of flavour symmetry is played by modular invariance.
The building blocks are modular forms of level $N$ and matter supermultiplets, both transforming in representations of a finite discrete group $\Gamma_N$. 
In the simplest version of these models, Yukawa couplings are just modular forms and the only source of flavour symmetry breaking is the vacuum expectation value of a single complex field, the modulus.
In the special case where modular forms are constant functions the whole construction collapses to a supersymmetric flavour model invariant under $\Gamma_N$, the case treated so far
in the literature. The framework has a number of appealing features. Flavon fields other than the modulus might not be needed.  Neutrino masses and 
mixing angles are simultaneously constrained by the modular symmetry. As long as supersymmetry is exact, modular invariance determines all higher-dimensional operators in the superpotential. 
We discuss the general framework and we provide complete examples 
of the new construction. The most economical model predicts neutrino mass ratios, lepton mixing angles, Dirac and Majorana phases uniquely in terms of the modulus vacuum expectation value, with all the parameters except one within the experimentally allowed range. 
As a byproduct of the general formalism we extend the notion of non-linearly realised symmetries to the discrete case.
\end{abstract}
\end{titlepage}
\setcounter{footnote}{0}
%%%%%%%%%%%%%%%%%%%%%%%%%%%%%%%%%%%%%%%%%%%%%%%%%%%%%%%%%%%%%%%%%%%%%%%%%%%%%%%%%%%%%%%%%%%
\thispagestyle{empty}
\begin{figure}[h!]
\centering
\includegraphics[width=0.8\textwidth]{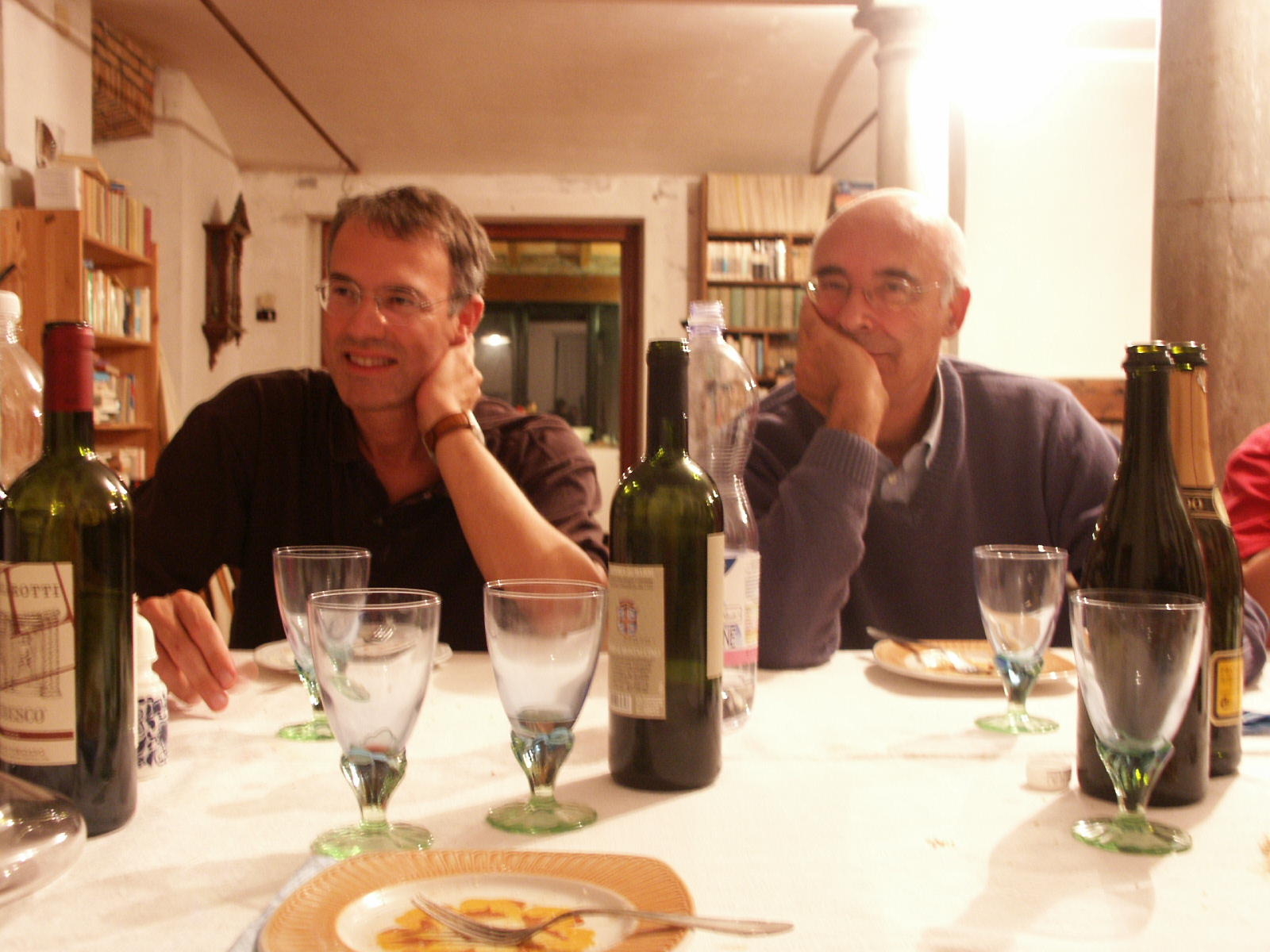}
%\caption{}
\end{figure}
\begin{center}
Parma, circa 2006
\end{center}
\vskip 2.cm
\vskip .5cm
\begin{center} 
\textit{ 
A contribution to:\\
``From my vast repertoire:  the legacy of Guido Altarelli"\\
S. Forte, A. Levy and G. Ridolfi, eds.}
\end{center}
\newpage
%%%%%%%%%%%%%%%%%%%%%%%%%%%%%%%%%%%%%%%%%%%%%%%%%%%%%%%%%%%%%%%%%%%%%%%%%%%%%%%%%%%%%%%%%%%
\section{Introduction}
%%%%%%%%%%%%%%%%%%%%%%%%%%%%%%%%%%%%%%%%%%%%%%%%%%%%%%%%%%%%%%%%%%%%%%%%%%%%%%%%%%%%%%%%%%%
A considerable interest in discrete flavour symmetries \cite{Altarelli:2010gt,Ishimori:2010au,Hernandez:2012ra,King:2013eh,King:2014nza,King:2017guk,Hagedorn:2017zks}
has been fostered by early models of quark masses and mixing angles \cite{Pakvasa:1977in,Wilczek:1977uh} and, more recently, by the discovery of neutrino oscillations.
Early data were well-compatible with a highly symmetric lepton mixing pattern, the tri-bimaximal one \cite{Harrison:2002er}, which could be derived from small non-abelian discrete symmetry groups such as 
$A_4$ \cite{Ma:2001dn,Babu:2002dz,Altarelli:2005yp}. Other discrete groups like $S_4$ and $A_5$ produced interesting alternative mixing patterns, which could be adopted as zeroth-order approximation to the data.
Today this approach is facing several difficulties. The formidable recent experimental progress has sharpened the neutrino
oscillation parameters, revealing many details that require a precise description, such as the non-vanishing value of the reactor angle, the deviation of the atmospheric angle
from the maximal value and a non-trivial Dirac CP-violating phase. Inclusion of these features in a realistic model based on discrete symmetries requires departure from minimality. Large corrections to the zeroth-order
approximation can be introduced at the price of spoiling predictability, due to the ignorance about the non-negligible higher-order contributions. Alternatively, groups of large dimensionality
can be invoked to correctly fit the data \cite{Toorop:2011jn,deAdelhartToorop:2011re,Holthausen:2012wt,King:2013vna,Hagedorn:2013nra,Fonseca:2014koa}. Discrete flavour symmetries can also be combined with CP invariance
in predictive models \cite{Feruglio:2012cw,Holthausen:2012dk}. Apart from the loss of minimality, there are several drawbacks in this program. The breaking of flavour symmetries typically relies on a generous set of scalar multiplets, the
so-called flavons,  and the Yukawa interactions generally include non-renormalizable operators with flavon insertions. Higher-dimensional operators with multiple flavon insertions come with unknown coefficients
that affect the model predictions. Moreover the flavon energy density has to be cleverly designed to get the correct vacuum alignment. The approach is mainly focused on 
lepton mixing angles while neutrino masses are reproduced by tuning the available parameters. Finally, it is not straightforward to extend the construction to the quark sector that seems not to like discrete symmetries.
In view of these disadvantages, anarchy \cite{Hall:1999sn,Haba:2000be,deGouvea:2003xe,Espinosa:2003qz,deGouvea:2012ac} 
and its generalizations have gained considerable momentum. Anarchy in the neutrino sector can arise in a variety of different frameworks
providing a  common description to both quark and lepton mass/mixing parameters,  also in the context of grand unified  theories \cite{Feruglio:2015jfa}.
However in the anarchy paradigm the observed lepton mixing angles
are regarded as environmental quantities \cite{Schellekens:2013bpa} and cannot be accurately predicted. For their intrinsic nature models based on anarchy essentially escape experimental tests aiming at an accuracy 
that matches the experimental precision.

In this wavering between order and anarchy we feel encouraged to investigate new directions.
Aim of the present work is to explore a new class of models generalizing the current approach based on discrete symmetry groups. 
These models are required to be invariant under transformations of the modular group, acting on the complex modulus $\tau$ (${\tt Im}(\tau)>0$) as linear fractional transformations: 
\be
\tau\to\frac{a \tau+b}{c \tau+d}~~~,~~~~~~~~~(a,b,c,d~{\rm integers}~,~~~~~ad-bc=1)~~~.\nn
\ee
In a supersymmetric theory these transformations naturally induce transformations of the matter multiplets according to representations of $\Gamma_N$, 
the so-called finite modular groups. Moreover there are holomorphic combinations of the modulus $\tau$, the 
modular forms of level $N$, that also transform in representations of $\Gamma_N$. We can thus exploit group theoretical properties of $\Gamma_N$ to build
supersymmetric, modular-invariant models of fermion masses, where matter fields and modular forms are used as building blocks.
A particular case of modular forms is that of constant functions. In this case the whole construction collapses to the case 
that have been considered in the literature so far when dealing with discrete flavour symmetries. On the other hand a new, presently unexplored, class of models arises when modular forms carry a non-trivial
dependence on $\tau$. For small $N$ the finite modular groups are
permutation groups and the new models are generalizations of well-known
existing models. 

In this paper we illustrate how the new construction can be realized, by discussing the general formalism
and by illustrating it with simple examples. In the minimal examples, see section 3.1.4, there are no flavons 
and the only source of flavour symmetry breaking is the vacuum expectation value (VEV) of the modulus $\tau$.
A truly remarkable feature of these examples is that all higher-dimensional operators in the superpotential are unambiguously determined in the limit of unbroken supersymmetry.
The possible corrections to the theory predictions can only come from the Kahler potential or from supersymmetry breaking contributions.
Moreover the modular symmetry constrains lepton mixing angles and neutrino masses at the same time. 
In the most economical examples, discussed in section 3.1.2 and 3.1.3, neutrino mass ratios, lepton mixing angles, Dirac and Majorana phases 
do not depend on any Lagrangian parameter, but only on the modulus VEV. In the non-trivial task of reproducing five known observables in terms of a single complex parameter,
the model of section 3.1.2 scores relatively well: only $\sin^2\theta_{13}$, predicted to be about $0.045$, is out of the allowed experimental range. The model predicts inverted neutrino mass ordering and determines completely the Majorana phases. The model of section 3.1.3 predicts normal neutrino mass ordering and is almost equally well-performing.

An unexpected byproduct of the new framework is a new concept of non-linearly realized symmetry, applied to discrete groups.
This is a built-in feature of modular forms of level $N$ but, in the second part of this work, we will analyze it in a more general framework.
In the context of relativistic quantum field theories, non-linear realizations of a continuous symmetry are synonymous of symmetry breaking and of 
the systematic low-energy expansion describing the massless modes predicted by the Goldstone theorem. The applications of these concepts are ubiquitous, 
ranging from the description of fundamental particle interactions to the theory of phase transitions. 
Extension of this formalism to the case of discrete symmetries may look unmanageable, especially when thinking to the Callan-Coleman-Wess-Zumino construction \cite{Coleman:1969sm,Callan:1969sn}, 
which is tightly connected to the Goldstone theorem. 
Moving from the properties of level-$3$ modular forms, we show how non-linear realizations of discrete symmetries can be defined and how they are related to the symmetry breaking pattern of 
a discrete group. In the continuous case non-linear realizations can be defined also by imposing invariant constraints
on the field space, such as for instance $\varphi^T \varphi=M^2$, the multiplet $\varphi$ transforming in the fundamental of $SO(N)$. 
In the discrete case this approach benefits from an important generalization. While a (connected, semisimple) Lie group has only the trivial one-dimensional
representation, a discrete group can possess a number of non-invariant singlets $\chi_i$. Therefore the invariant conditions $\chi_i=0$ have no counterpart in 
(connected, semisimple) Lie groups. If the product of a set of multiplets $\varphi$ contains one ore more non-invariant singlets $\chi_i(\varphi)$, the conditions
 $\chi_i(\varphi)=0$ may define a non-linear realization of the discrete group. This is precisely what happens with modular forms of level 3. Non-linear realizations of a discrete symmetry do not define low-energy effective
 theories. They should rather be view as consistent truncations of some ultraviolet completion. As we will see, in general, the requirement  $\chi_i(\varphi)=0$ leaves no non-trivial
 residual symmetry. We identify the cases in which a non-trivial residual symmetry group survives and we discuss the relation of such event with the properties of 
 the so-called orbit space, the space spanned by the invariants of the group.  This last discussion is only very tentative,
 but we hope that in the future it may be better clarified, offering further possibilities in model building.
%%%%%%%%%%%%%%%%%%%%%%%%%%%%%%%%%%%%%%%%%%%%%%%%%%%%%%%%%%%%%%%%%%%%%%%%%%%%%%%%%%%%%%%%%%%
\section{Modular Forms and Modular Invariant Theories}
%%%%%%%%%%%%%%%%%%%%%%%%%%%%%%%%%%%%%%%%%%%%%%%%%%%%%%%%%%%%%%%%%%%%%%%%%%%%%%%%%%%%%%%%%%%
In this section we briefly recap definitions and properties of modular forms. We also review the framework of modular invariant supersymmetric theories. The material of this Section is quite known,
but will be presented here from a different perspective, emphasizing the role of finite discrete symmetries in modular invariant theories. 
Modular invariance has a long history both in string and in field theories.
Target space modular invariance has been investigated soon after the discovery that the spectrum of a closed string, 
when compactified on a circle of radius $R$, is invariant under the duality transformation $R\to 1/2R$, which corresponds to the linear fractional transformation $\tau\to -1/\tau$ restricted to 
the imaginary part of $\tau$ ($\tau=2 i R^2$). Modular invariance controls orbifold compactifications of the heterotic string and requires that the
couplings among twisted states are modular forms \cite{Hamidi:1986vh,Dixon:1986qv,Lauer:1989ax,Lauer:1990tm,Erler:1991nr}.  This holds in particular for realistic Yukawa couplings \cite{Ibanez:1986ka,Casas:1991ac,Lebedev:2001qg,Kobayashi:2003vi}.
In orientifold compactifications of Type II strings the Yukawa couplings are functions with specific modular properties \cite{Cremades:2003qj,Blumenhagen:2005mu,Abel:2006yk,Blumenhagen:2006ci,Marchesano:2007de,Antoniadis:2009bg,Kobayashi:2016ovu}.
A similar feature is found also in magnetised extra dimensions \cite{Cremades:2004wa,Abe:2009vi}.
Modular invariant supersymmetric field theories have been analyzed in the late 80s \cite{Ferrara:1989bc,Ferrara:1989qb}, both for global and local supersymmetry.
Modular invariance in field theory constructions has been invoked while addressing several aspects of the flavour problem in model building
\cite{Brax:1994kv,Binetruy:1995nt,Dudas:1995eq,Dudas:1996aa,Leontaris:1997vw,Dent:2001cc,Dent:2001mn}.
Duality and modular invariance have been  suggested as underlying properties of the quantum Hall effect \cite{Lutken:1991jk,Cappelli:1996np,Dolan:1998vs,Dolan:1998vr,Burgess:2000kj,Burgess:2001sy,Burgess:2006fw,Lutken:2007zz,Lutken:2011zz}.

%%%%%%%%%%%%%%%%%%%%%%%%%%%%%%%%%%%%%%%%%%%%%%%%%%%%%%%%%%%%%%%%%%%%%%%%%%%%%%%%%%%%%%%%%%%
\subsection{Modular forms of level $N$}
%%%%%%%%%%%%%%%%%%%%%%%%%%%%%%%%%%%%%%%%%%%%%%%%%%%%%%%%%%%%%%%%%%%%%%%%%%%%%%%%%%%%%%%%%%%
We consider the series of groups $\Gamma(N)$ $(N=1,2,3,....)$ defined by:
\be
\Gamma(N)=\{\left(\begin{array}{cc}a&b\\c&d\end{array}\right)\in SL(2,Z), \left(\begin{array}{cc}a&b\\c&d\end{array}\right)=\left(\begin{array}{cc}1&0\\0&1\end{array}\right)~~~({\tt mod}~N)\}~~~,
\ee
where $\Gamma\equiv SL(2,Z)$ is the group of two by two matrices with integer entries and determinant equal to one, also called homogeneous modular group. We have $\Gamma=\Gamma(1)$ and the groups $\Gamma(N)$ $(N\ge 2)$ 
are infinite normal subgroups of $\Gamma$, called principal congruence subgroups.
The group $\Gamma(N)$ acts on the complex variable $\tau$, varying in the upper-half complex plane ${\cal H}={\tt Im}(\tau)>0$, as the linear fractional transformation
\be
\gamma \tau=\frac{a \tau+b}{c \tau+d}~~~.
\label{lft}
\ee
The quotient space ${\cal H}/\Gamma(N)$ can be compactified by adding special points called cusps. They coincide with $i\infty$ or with rational real numbers.
While it is very convenient to deal with two by two matrices, the groups $\overline{\Gamma}(N)$ of linear fractional transformations are slightly different from the groups $\Gamma(N)$.
Since $\gamma$ and $-\gamma$ induce the same linear fractional transformation, for $N=1$ and $N=2$, the transformations (\ref{lft}) are in a one-to-one correspondence
with the elements of the group $\overline{\Gamma}\equiv \Gamma/\{\pm  \mathds{1}\}$ and $\overline{\Gamma}(2)\equiv \Gamma(2)/\{\pm  \mathds{1}\}$,
respectively.  $\overline{\Gamma}$ is called inomhogeneous modular group, or simply modular group. When $N>2$ the element $-\mathds{1}$ does not belong to $\Gamma(N)$ and we have $\overline{\Gamma}(N)\equiv \Gamma(N)$.
The group $\overline{\Gamma}$ is generated by two elements $S$ and $T$ satisfying:
\be
S^2=\mathds{1}~~,~~(ST)^3=\mathds{1}~~~.
\label{modular}
\ee
They  can be represented by  the $SL(2,Z)$ matrices:
\be
S=\left(
\begin{array}{cc}
0&1\\
-1&0
\end{array}
\right)~~~~~~~~~~~~
T=\left(
\begin{array}{cc}
1&1\\
0&1
\end{array}
\right)~~~,
\label{sandt}
\ee
corresponding to the transformations
\be
S: \tau \rightarrow -\frac{1}{\tau}~~~,~~~~~~~~
T: \tau \rightarrow \tau+1~~~.
\ee
The quotient groups  $\Gamma_N\equiv \overline{\Gamma}/\overline{\Gamma}(N)$ are called finite modular groups. Some of their properties can be found in ref. \cite{deAdelhartToorop:2011re}.

Modular forms $f(\tau)$ of weight $2k$ and level $N$ are holomorphic functions of the complex variable $\tau$ 
with well-defined transformation properties under the group $\Gamma(N)$ \cite{gunning}:
\be
f(\gamma \tau)=(c \tau+d)^{2k} f(\tau)~~~~~~~~~~\gamma=\left(\begin{array}{cc}a&b\\c&d\end{array}\right)\in \Gamma(N)~~~,
\ee
where $k\ge 0$ is an integer \footnote{Following \cite{gunning}, here we only consider modular forms of even weight.}. The function $f(\tau)$ is required to be holomorphic in ${\cal H}$ and at all the cusps.
In the special case when it vanishes at all the cusp, $f(\tau)$ is called a cusp form.
For $\Gamma(N)$ $(N\ge 2)$, $N$ is the level of the group and $T^N$ belongs to $\Gamma(N)$. It follows that $f(\tau+N)=f(\tau)$ and we have the Fourier expansion (or simply $q$-expansion):
\be
f(\tau)=\sum_{i=0}^\infty a_n q_N^n~~~~~~~~~q_N=e^{\frac{i 2\pi \tau}{N}}~~~.
\ee
For $k<0$ there are no modular forms.
If $k=0$ the only possible modular form is a constant. Modular forms of weight $2k$ and level $N$
form a linear space ${\cal M}_{2k}(\Gamma(N))$ of finite dimension $d_{2k}(\Gamma(N))$. The dimensions $d_{2k}(\Gamma(N))$ for few levels $N$ are shown in table 1, derived  from ref. \cite{gunning}. General formulas are reported in Appendix A.
\begin{table}[h] 
\centering
\begin{tabular}{|c|c|c|c|c|}
\hline
$N$&$g$ &$d_{2k}(\Gamma(N))$&$\mu_N$& $\Gamma_N$\rule[-2ex]{0pt}{5ex}\\
\hline
2&0&$k+1$&6&$S_3$\rule[-2ex]{0pt}{5ex}\\
\hline
3&0&$2k+1$&12&$A_4$\rule[-2ex]{0pt}{5ex}\\
\hline
4&0&$4k+1$&24&$S_4$\rule[-2ex]{0pt}{5ex}\\
\hline
5&0&$10k+1$&60&$A_5$\rule[-2ex]{0pt}{5ex}\\
\hline
6&1&$12k$&72&\rule[-2ex]{0pt}{5ex}\\
\hline
7&3&$28k-2$&168&\rule[-2ex]{0pt}{5ex}\\
\hline
\end{tabular}
\caption{Some properties of modular forms: $g$ is the genus of the space ${\cal H}/\Gamma(N)$ after compactification, $d_{2k}(\Gamma(N))$ the dimension of the linear space ${\cal M}_{2k}(\Gamma(N))$, $\mu_N$ is the dimension of the quotient group
$\Gamma_N\equiv \overline{\Gamma}/\overline{\Gamma}(N)$, which, for $N\le 5$, is isomorphic to a permutation group.}
\label{tabmod}
\end{table}
The product of two modular forms of level $N$ and weights $2k$, $2k'$ is a modular form of level $N$ and weight $2(k+k')$ and the set ${\cal M}(\Gamma(N))$ of all modular forms of level $N$
is a ring
\be
{\cal M}(\Gamma(N))=\bigoplus_{k=0}^{\infty}{\cal M}_{2k}(\Gamma(N))~~~,
\ee
generated by few elements. For instance ${\cal M}(\Gamma)$ is generated by two modular forms $E_4(\tau)$ and $E_6(\tau)$ of weight 4 and 6 respectively,
so that each modular form in ${\cal M}_{2k}(\Gamma)$ can be written as a polynomial $\sum_{ij} c_{ij}~ E_4(\tau)^{n_i} E_6(\tau)^{n_j}$, with powers satisfying $2k=4 n_i+6 n_j$.

Central to the construction illustrated in the next Sections is the following result, explicitly proved in Appendix B. Modular forms of weight $2k$ and level $N\ge 2$ are invariant, up to the factor $(c\tau+d)^{2k}$, under $\Gamma(N)$ but they transform under the quotient group $\Gamma_N\equiv \overline{\Gamma}/\overline{\Gamma}(N)$. As show in Appendix B, it is possible to choose a basis in ${\cal M}_{2k}(\Gamma(N))$ such that this transformation is described by a unitary representation $\rho$ of $\Gamma_N$:
\be
f_i(\gamma \tau)=(c \tau+d)^k \rho(\gamma)_{ij}f_j(\tau)~~~,
\ee
where
\be
\gamma=\left(\begin{array}{cc}a&b\\c&d\end{array}\right)
\ee
stands for a representative element of $\Gamma_N$. In practise, it is sufficient to determine the representation $\rho$ for the two elements $S$ and $T$ that generate the entire modular group
and, for $N\ge 2$, are not contained in $\overline{\Gamma}(N)$. While the groups $\overline{\Gamma}(N)$ are infinite, the quotient groups $\Gamma_N$ are finite. In table \ref{tabmod} we show few groups $\Gamma_N$ and their
dimensions. 

%%%%%%%%%%%%%%%%%%%%%%%%%%%%%%%%%%%%%%%%%%%%%%%%%%%%%%%%%%%%%%%%%%%%%%%%%%%%%%%%%%%%%%%%%%%
\subsection{Modular-invariant supersymmetric theories}
%%%%%%%%%%%%%%%%%%%%%%%%%%%%%%%%%%%%%%%%%%%%%%%%%%%%%%%%%%%%%%%%%%%%%%%%%%%%%%%%%%%%%%%%%%%
Here we briefly review the formalism \cite{Ferrara:1989bc,Ferrara:1989qb}, starting from the case of $N=1$ global supersymmetry, where the action takes the general form\footnote{We focus on Yukawa interactions and we turn off the gauge interactions.}:
\be
{\cal S}=\int d^4 x d^2\theta d^2\bar \theta~ K(\Phi,\bar \Phi)+\int d^4 x d^2\theta~ w(\Phi)+h.c.
\label{action}
\ee
where $K(\Phi,\bar\Phi)$, the Kahler potential, is a real gauge-invariant function of the chiral superfields $\Phi$ and their conjugates
and $w(\Phi)$, the superpotential, is a holomorphic gauge-invariant function of the chiral superfields $\Phi$.
With $\Phi=(\tau,\varphi)$ we denote collectively all chiral supermultiplets of the theory, including the modulus \footnote{In the literature it is common to make use of the field $T=-i\tau$. The modulus $\tau$ describes a dimensionless chiral supermultiplet, depending on both space-time and Grassmann coordinates. By introducing a fundamental scale $\Lambda$, a chiral supermultiplet $\sigma=\Lambda\tau$ with mass dimension one can be defined.
In the limit of unbroken supersymmetry the results presented in this paper depend on $\sigma$ only through the $\tau$ combination.}  $\tau$ plus a number of additional chiral supermultiplets $\varphi$ separated into sectors $\varphi^{(I)}$. In general each sector $I$ contains
several chiral supermultiplets $\varphi^{(I)}_{i_I}$ but, to keep our notation more concise, we will not write the additional index $i_I$.
We ask for invariance under transformations of the modular group $\overline{\Gamma}$, under which the supermultiplets $\varphi^{(I)}$ of each sector $I$ are assumed to transform in a representation $\rho^{(I)}$ of a quotient group $\Gamma_N$,  kept fixed in the construction, with a weight $-k_I$:
\be
\left\{
\begin{array}{l}
\tau\to \dd\frac{a \tau+b}{c \tau+d}\\[0.2 cm]
\varphi^{(I)}\to (c\tau+d)^{-k_I} \rho^{(I)}(\gamma) \varphi^{(I)}
\end{array}~~~.
\right.
\label{tmg}
\ee
The supermultiplets $\varphi^{(I)}$ are not modular forms and there are no restrictions on the possible value of $k_I$, a priori.
The invariance of the action ${\cal S}$ under (\ref{tmg}) 
requires the invariance of the superpotential $w(\Phi)$ and the invariance of the Kahler potential up to a Kahler transformation:
\be
\left\{
\begin{array}{l}
w(\Phi)\to w(\Phi)\\[0.2 cm]
K(\Phi,\bar \Phi)\to K(\Phi,\bar \Phi)+f(\Phi)+f(\overline{\Phi})
\end{array}
\right.~~~.
\ee
The requirement of invariance of the Kahler potential can be easily satisfied. An example of Kahler potential invariant under (\ref{tmg}) up to Kahler transformations is
\be
K(\Phi,\bar \Phi)=-h \log(-i\tau+i\bar\tau)+ \sum_I (-i\tau+i\bar\tau)^{-k_I} |\varphi^{(I)}|^2~~~,
\label{kalex}
\ee
where $h$ is a positive constant.
Since $(\tau-\bar\tau)\to |c \tau+d|^{-2}(\tau-\bar\tau)$, we have:
\be
K(\Phi,\bar \Phi)\to K(\Phi,\bar \Phi)+h \log(c \tau+d)+h \log(c\bar\tau+d)~~~.
\ee
Assuming that only the modulus $\tau$ acquires a vacuum expectation value (VEV), this Kahler potential gives rise to kinetic terms for the scalar components of the supermultiplets $\tau$ and $\varphi^{(I)}$ of the kind
\be
\frac{h}{\langle-i\tau+i\bar\tau\rangle^2}\partial_\mu \bar\tau\partial^\mu \tau+\sum_I \frac{\partial_\mu  \overline{\varphi}^{(I)}\partial^\mu \varphi^{(I)}}{\langle-i\tau+i\bar\tau\rangle^{k_I}}~~~.
\ee
On the contrary, the invariance of the superpotential $w(\Phi)$ under the modular group provides a strong restriction on the theory. Consider the expansion of $w(\Phi)$ in power series of the supermultiplets $\varphi^{(I)}$:
\be
w(\Phi)=\sum_n Y_{I_1...I_n}(\tau)~ \varphi^{(I_1)}... \varphi^{(I_n)}~~~.
\label{psex}
\ee
For the $n$-th order term to be modular invariant the functions $Y_{I_1...I_n}(\tau)$ should be modular forms of weight $k_Y(n)$ transforming
in the representation $\rho$ of $\Gamma_N$:
\be
Y_{I_1...I_n}(\gamma\tau)=(c\tau+d)^{k_Y(n)} \rho(\gamma)~Y_{I_1...I_n}(\tau)~~~,
\ee
with $k_Y(n)$ and $\rho$ such that:
\begin{enumerate}
\item[1.]
The weight $k_Y(n)$ should compensate the overall weight of the product $\varphi^{(I_1)}... \varphi^{(I_n)}$:
\be
k_Y(n)=k_{I_1}+....+k_{I_n}~~~.
\label{compensate}
\ee
\item[2.]
The product $\rho\times \rho^{{I_1}}\times ... \times \rho^{{I_n}}$ contains an invariant singlet.
\end{enumerate}
When we have $k_I=0$ in all sectors $I$ of the theory, we get $k_Y(n)=0$ for all $n$ and the functions $Y_{I_1...I_n}(\tau)$ are $\tau$-independent constants
since, as we have seen, there are no non-trivial modular forms of weight zero.
The product $\rho^{{I_1}}\times ... \times \rho^{{I_n}}$ should contain an invariant singlet and $Y_{I_1...I_n}$ is an invariant tensor under the group $\Gamma_N$.
This particular limit reproduces the well-studied case of a supersymmetric theory invariant under the discrete symmetry $\Gamma_N$. Some of the fields $\varphi^{(I)}$ can be
gauge singlets, {\em i.e.} flavons, whose vacuum expectation values break $\Gamma_N$ in the appropriate direction.  This scheme applies to
most of the models of fermion masses based on discrete symmetries that have been proposed so far.
However this is just a particular case of the more general setup considered here.
When $k_I\ne 0$ in some sector $I$, the Yukawa couplings $Y_{I_1...I_n}(\tau)$ should carry a non-trivial $\tau$ dependence. The whole modular group acts
on the field space and $Y_{I_1...I_n}(\tau)$ are strictly constrained by the relatively small number of possible modular forms that match the above requirements. 

This setup can be easily extended to the case of $N=1$ local supersymmetry where Kahler potential and superpotential are not independent
functions since the theory depends on the combination
\be
{\cal G}(\Phi,\bar\Phi)=K(\Phi,\bar \Phi)+\log w(\Phi)+\log w(\bar\Phi)~~~.
\ee
The modular invariance of the theory can be realized in two ways. Either $K(\Phi,\bar \Phi)$ and $w(\Phi)$ are separately modular invariant or
the transformation of $K(\Phi,\bar \Phi)$ under the modular group is compensated by that of $w(\Phi)$. An example of this second possibility is given
by the Kahler potential of eq. (\ref{kalex}), with the superpotential $w(\Phi)$ transforming as
\be
w(\Phi)\to e^{i\alpha(\gamma)} (c\tau+d)^{-h} w(\Phi) 
\ee
In the expansion (\ref{psex}) the Yukawa couplings $Y_{I_1...I_n}(\tau)$ should now transform as 
\be
Y_{I_1...I_n}(\gamma\tau)= e^{i\alpha(\gamma)} (c\tau+d)^{k_Y(n)} \rho(\gamma)~Y_{I_1...I_n}(\tau)~~~,
\ee
with $k_Y(n)=k_{I_1}+....+k_{I_n}-h$ and the representation $\rho$ subject to the requirement 2.  
When we have $k_{I_1}+....+k_{I_n}=h$, we get $k_Y(n)=0$ and the functions $Y_{I_1...I_n}(\tau)$ are $\tau$-independent constants.
This occurs for supermultiplets belonging to the untwisted sector in the orbifold compactification of the heterotic string.

The above framework is known and we have reported it here for completeness. However there are two aspects that we have emphasized in this presentation. First of all, the close relationship between supersymmetric modular-invariant theories and finite discrete symmetries. As stressed above the whole class of models
invariant under discrete groups of the type $\Gamma_N$ can be regarded as a particular case of the wider class of modular invariant theories. The second aspect
is that this formulation is particular suitable to a bottom-up approach.
Namely, by explicitly constructing the whole ring ${\cal M}(\Gamma(N))$ of modular forms of level $N$, one can systematically explore all possible Yukawa couplings
$Y_{I_1...I_n}(\tau)$ which can occur in the superpotential $w(\Phi)$. As we shall see in a specific example, the limited number of generators of ${\cal M}(\Gamma(N))$
severely constrain the candidate couplings and lead to a new class of predictive models of lepton masses. 
%%%%%%%%%%%%%%%%%%%%%%%%%%%%%%%%%%%%%%%%%%%%%%%%%%%%%%%%%%%%%%%%%%%%%%%%%%%%%%%%%%%%%%%%%%%
\section{Models of Lepton Masses }
%%%%%%%%%%%%%%%%%%%%%%%%%%%%%%%%%%%%%%%%%%%%%%%%%%%%%%%%%%%%%%%%%%%%%%%%%%%%%%%%%%%%%%%%%%%
In this section we apply the formalism outlined above to models of lepton masses and mixings. Here we give the general rules to build
modular invariant $N=1$ supersymmetric models with matter supermultiplets transforming according to representations of a finite modular group
$\Gamma_N$, with a fixed integer $N$. The relevant matter fields and their transformation properties are collected in table \ref{tabmod1}. \begin{table}[h] 
\centering
\begin{tabular}{|c|c|c|c|c|c|c|}
\hline
&$E^c$& $N^c$& $L$& $H_d$& $H_u$& $\varphi$\rule[-2ex]{0pt}{5ex}\\
\hline
$SU(2)_L\times U(1)_Y$&$(1,+1)$& $(1,0)$& $(2,-1/2)$& $(2,-1/2)$& $(2,+1/2)$& $(1,0)$ \rule[-2ex]{0pt}{5ex}\\
\hline
$\Gamma_N$&$\rho_E$& $\rho_N$& $\rho_L$& $\rho_d$& $\rho_u$ &$\rho_\varphi$ \rule[-2ex]{0pt}{5ex}\\
\hline
$k_I$&$k_{E_i}$& $k_{N_i}$& $k_{L_i}$& $k_d$& $k_u$& $k_{\varphi_i}$ \rule[-2ex]{0pt}{5ex}\\
\hline
\end{tabular}
\caption{Chiral supermultiplets, transformation properties and weights. Matter supermultiplets $E^c_i$, $N^c_i$ and $L_i$ come in three copies and their representations $\rho_I$ 
$(I=E,N,L)$ can be reducible. If this is the case,  different generations can have different weight. $\varphi$ denotes the flavons, a set of gauge-invariant chiral supermultiplets that can contribute
to the spontaneous breaking of $\Gamma_N$ through a non-vanishing VEV.}
\label{tabmod1}
\end{table}
The field content includes a chiral supermultiplet for  the modulus $\tau$, transforming as in (\ref{tmg}) under modular transformations and singlet under gauge transformations. 
There is a minimal class of models where the only source of breaking of the modular symmetry is the VEV of $\tau$. In this class of models we do not introduce
any flavon multiplet. The role of flavons is replaced by the modular forms $Y(\tau)$ of level $N$. In a more general context, we can allow for the existence of a set of gauge-singlet flavons $\varphi$,
which transform non-trivially under $\Gamma_N$ and contribute to the breaking of the discrete symmetry through their VEVs. Without loosing generality the flavon fields can be
assumed dimensionless, as the modulus $\tau$.
As a preliminary step, we need to identify the modular forms $Y(\tau)$ and their transformation properties under the modular group, {\em i.e.} their representations $\rho$ under $\Gamma_N$.
We assume a Kahler potential of the form (\ref{kalex}). For the superpotential, we distinguish two cases.
\begin{itemize}
\item[1.]
When neutrino masses originate from a type I see-saw mechanism,  the superpotential in the lepton sector reads:
\be
w=\alpha~ (E^c H_d L~ f_E(Y,\varphi))_1+g(N^c H_u L~ f_N(Y,\varphi))_1+\Lambda (N^c N^c f_M(Y,\varphi))_1~~~,
\label{su1}
\ee
where $f_I(Y,\varphi)$ $(I=E,N,M)$ denotes the most general combination of  modular forms $Y(\tau)$ and flavon fields $\varphi$ such that the total weight of each term in (\ref{su1}) is zero.
Moreover $(...)_1$ stands for the invariant singlet of $\Gamma_N$. It can occur that there are several independent invariant singlets. 
In this case a sum over all contributions with arbitrary coefficients is understood and in $w$ there are more parameters than those explicitly indicated.
\item[2.]
When neutrino masses directly originate from the Weinberg operator,  the superpotential in the lepton sector reads:
\be
w=\alpha~ (E^c H_d L~ f_E(Y,\varphi))_1+\dd\frac{1}{\Lambda}(H_u H_u~ L L~ f_W(Y,\varphi))_1~~~.
\label{su1}
\ee
\end{itemize}
The supermultiplet $\tau$ and the modular forms $Y(\tau)$ do not carry lepton number. In the above superpotential the lepton number of the matter multiplets is assigned
in the usual way and the total lepton number is effectively broken either by the VEV of some flavon field or by the parameter $\Lambda$, that can be regarded as a spurion.
We will not address here the vacuum alignment problem. We will not attempt to build the most general supersymmetric and modular invariant scalar potential for $\tau$ and $\varphi$.
To explore the ability of this class of models in reproducing the existing data, both $\tau$ and $\varphi$ will be treated as spurions whose valued will be varied when scanning the parameter
space of the model. In principle, for each value of $\tau$ the kinetic terms of the chiral multiplets in table \ref{tabmod1} have to be rescaled to match their canonical form.
In practice, in the concrete models we will explore, this effect can be absorbed into the unknown parameters of the superpotential.
By scanning the parameter space of the model $(\alpha,g,\Lambda)$ and the VEVs $(\langle\tau\rangle,\langle\varphi\rangle)$ we can compute lepton masses, mixing angles and phases and compare them to the data.
%%%%%%%%%%%%%%%%%%%%%%%%%%%%%%%%%%%%%%%%%%%%%%%%%%%%%%%%%%%%%%%%%%%%%%%%%%%%%%%%%%%%%%%%%%%
\subsection{Models based on $\Gamma_3$ }
%%%%%%%%%%%%%%%%%%%%%%%%%%%%%%%%%%%%%%%%%%%%%%%%%%%%%%%%%%%%%%%%%%%%%%%%%%%%%%%%%%%%%%%%%%%
To assess the viability of the above formalism, we build in this section models of lepton masses based on modular forms of level 3. The group $\Gamma_3$ is isomorphic to $A_4$,
see Table 1, and we will be able to compare the present construction with a class of models widely discussed in the literature, especially in connection to the tribimaximal mixing.
Though the explicit example will refer to the case $N=3$, it is pretty  clear from the previous section that there are no conceptual obstacle to develop a similar construction for any $N$.
A comprehensive analysis of all models that can be constructed along these lines is beyond the scope of the present work. The examples discussed here, admittedly not fully realistic, are meant
to illustrate how this new type of construction can be brought to completion and compared with the data.
%%%%%%%%%%%%%%%%%%%%%%%%%%%%%%%%%%%%%%%%%%%%%%%%%%%%%%%%%%%%%%%%%%%%%%%%%%%%%%%%%%%%%%%%%%%
\subsubsection{Modular Forms of Level 3 }
%%%%%%%%%%%%%%%%%%%%%%%%%%%%%%%%%%%%%%%%%%%%%%%%%%%%%%%%%%%%%%%%%%%%%%%%%%%%%%%%%%%%%%%%%%%
We focus on modular forms of level 3, satisfying
\be
f(\gamma \tau)=(c \tau+d)^{2k} f(\tau)~~~~~~~~~~\gamma=\left(\begin{array}{cc}a&b\\c&d\end{array}\right)\in \Gamma(3)~~~,
\ee
with
\be
\Gamma(3)=\{\left(\begin{array}{cc}a&b\\c&d\end{array}\right)\in SL(2,Z), \left(\begin{array}{cc}a&b\\c&d\end{array}\right)=\left(\begin{array}{cc}1&0\\0&1\end{array}\right)~~~({\tt mod}~3)\}~~~.
\ee
The quotient space ${\cal H}/\Gamma(3)$ can be described by a fundamental domain ${\cal F}$ for $\Gamma(3)$, that is a connected region of ${\cal H}$ such that
each point of ${\cal H}$ can be mapped into ${\cal F}$ by a $\Gamma(3)$ transformation, but no two points in the interior of ${\cal F}$ are related under $\Gamma(3)$.
The space ${\cal H}/\Gamma(3)$ is simply ${\cal F}$ with certain boundary points identified.
A fundamental domain for $\Gamma(3)$ is shown in fig. 1.
\begin{figure}[h!]
\centering
\includegraphics[width=0.5\textwidth]{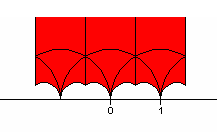}
\caption{Fundamental domain for $\Gamma(3)$.}
\end{figure}
${\cal H}/\Gamma(3)$ can be made compact by adding the points $i\infty$, $-1$, $0$ and $+1$, which are the cusps.
The compactified space $\overline{{\cal H}/\Gamma(3)}$ has genus zero and can be thought of as a tetrahedron 
whose vertices are the cusps. Indeed the cusps  are related by transformations of $\Gamma_3=\bar\Gamma/\overline{\Gamma}(3)$, which is the symmetry group of a regular tetrahedron,
given the isomorphism between $\Gamma_3$ and $A_4$. The group $A_4$ is generated by the elements $S$ and $T$, satisfying the relations:
\be
S^2=T^3=(ST)^3=\mathds{1}~~~.
\ee
Basic properties of $A_4$ are summarized in Appendix C. 
Modular forms of weight $2k$ and level $3$ transform according to unitary representations of $A_4$. In the final part of this work we will see
that they provide a non-linear representation of $A_4$. Modular forms of level 3 can be constructed starting from those of lower weight, $k=1$. From table 1 we see that there are three linearly independent such forms, which we call $Y_i(\tau)$. Three linearly independent weight 2 and level-3 forms
are constructed in the Appendix C. They read:
\bea
\label{craft}
Y_1(\tau)&=&\frac{i}{2 \pi}
   \left[
   \frac{\eta'\left(\frac{\tau }{3}\right)}{\eta \left(\frac{\tau}{3}\right)}
   +\frac{\eta'\left(\frac{\tau+1}{3}\right)}{\eta \left(\frac{\tau+1}{3}\right)}
   +\frac{\eta'\left(\frac{\tau +2}{3}\right)}{\eta\left(\frac{\tau +2}{3}\right)}
   -\frac{27 \eta'(3 \tau )}{\eta (3 \tau)}
   \right]\nn\\
Y_2(\tau)&=&\frac{-i}{\pi}
   \left[
   \frac{\eta'\left(\frac{\tau }{3}\right)}{\eta \left(\frac{\tau}{3}\right)}
+\omega^2~\frac{\eta'\left(\frac{\tau+1}{3}\right)}{\eta \left(\frac{\tau+1}{3}\right)}
   +\omega~\frac{\eta'\left(\frac{\tau +2}{3}\right)}{\eta\left(\frac{\tau +2}{3}\right)}
   \right]\\
Y_2(\tau)&=&\frac{-i}{\pi}
   \left[
   \frac{\eta'\left(\frac{\tau }{3}\right)}{\eta \left(\frac{\tau}{3}\right)}
+\omega~\frac{\eta'\left(\frac{\tau+1}{3}\right)}{\eta \left(\frac{\tau+1}{3}\right)}
   +\omega^2~\frac{\eta'\left(\frac{\tau +2}{3}\right)}{\eta\left(\frac{\tau +2}{3}\right)}
   \right]~~~.\nn
\eea
where $\eta(\tau)$ is the Dedekind eta-function, defined in the upper complex plane:
\be
\eta(\tau)=q^{1/24}\prod_{n=1}^\infty \left(1-q^n \right)~~~~~~~~~~~~~q\equiv e^{i 2 \pi\tau}~~~.
\ee
They transform in the three-dimensional representation of $A_4$. In a vector notation where $Y^T=(Y_1,Y_2,Y_3)$ we have
\be
Y(-1/\tau)=\tau^2~\rho(S) Y(\tau)~~~,~~~~~~~~~~Y(\tau+1)=\rho(T) Y(\tau)~~~,
\nn
\ee
with unitary matrices
$\rho(S)$ and $\rho(T)$
\vskip 0.1cm
\be
\rho(S)=\frac{1}{3}
\left(
\begin{array}{ccc}
-1&2&2\\
2&-1&2\\
2&2&-1
\end{array}
\right)~~~,~~~~~~~~~~
\rho(T)=
\left(
\begin{array}{ccc}
1&0&0\\
0&\omega&0\\
0&0&\omega^2
\end{array}
\right)~~~,~~~~~~~\omega=-\frac{1}{2}+\frac{\sqrt{3}}{2}i~~~.
\nn
\ee
\vskip 0.1cm
\noindent
The $q$-expansion of $Y_i(\tau)$ reads:
\bea
Y_1(\tau)&=&1+12q+36q^2+12q^3+...\nn\\
Y_2(\tau)&=&-6q^{1/3}(1+7q+8q^2+...)\nn\\
Y_3(\tau)&=&-18q^{2/3}(1+2q+5q^2+...)~~~.\nn
\eea
From the $q$-expansion we see that the functions $Y_i(\tau)$ are regular at the cusps.
Moreover $Y_i(\tau)$ satisfy the constraint:
\be
Y_2^2+2 Y_1 Y_3=0~~~.
\label{zmf}
\ee
As discussed explicitly in Appendix D,
the constraint (\ref{zmf}) is essential to recover the correct dimension of the linear space ${\cal M}_{2k}(\Gamma(3))$.
On the one side from table 1 we see that this space has dimension $2k+1$. On the other hand the number of independent homogeneous polynomial $Y_{i_1}Y_{i_2} \cdot\cdot\cdot Y_{i_k}$ of degree 
$k$ that we can form with $Y_i$ is $(k+1)(k+2)/2$. These polynomials are modular forms of weight $2k$ and, to match the correct dimension,
$k(k-1)/2$ among them  should vanish. Indeed this happens as a consequence of eq. (\ref{zmf}).
Therefore the ring ${\cal M}(\Gamma(3))$ is generated by the modular forms $Y_i(\tau)$ $(i=1,2,3)$.

There are two special sets of VEV for $\tau$ that preserve a subgroup of the modular group (and of $\Gamma_3$). 
The subgroup generated by $T$ is preserved by $\langle \tau\rangle=i\infty$
\footnote{This can be seen more clearly by going from ${\cal H}$ to the unit disk $|w|<1$, via $w=(\tau-1)/(\tau+1).$ The transformation $T$ acts on $w$ as $w\to(1+w)/(3-w)$, having a unique fixed point at
$w=1$, belonging to the boundary of the unit disk and corresponding to $\tau=i\infty$.}. This induces the well-known VEV for the triplet $Y_i$:
\be
(Y_1,Y_2,Y_3)|_{\tau=i\infty}=(1,0,0)~~~.
\ee
The subgroup generated by $S$ is preserved by $\langle \tau\rangle=i$, which results into:
\be
(Y_1,Y_2,Y_3)|_{\tau=i}=Y_1(i)(1,1-\sqrt{3},-2+\sqrt{3})
\label{buono}
\ee
This VEV pattern is completely different from $(1,1,1)$, the eigenvector of the matrix $S$ corresponding to the eigenvalue +1. Indeed we should satisfy:
\be
Y(-1/\tau)|_{\tau=i}=-\rho(S) Y(\tau)|_{\tau=i}~~~.
\ee
For this reason $Y(\tau)|_{\tau=i}$ must be an eigenvector of the matrix $S$ with eigenvalue -1. There are two such eigenvectors and the one in (\ref{buono})
satisfies the relation (\ref{zmf}). Notice that the configuration $(Y_1,Y_2,Y_3)=(1,1,1)$, widely used in model building, cannot be realised since it violates the constraint (\ref{zmf}).

Of course, values of $\langle\tau\rangle$ related to $i$ by a modular transformation, have little groups
isomorphic to $Z_2$, the subgroup generated by $S$. A similar consideration applies to the vacuum configurations related to $\langle\tau\rangle=i\infty$ by modular transformations.
%%%%%%%%%%%%%%%%%%%%%%%%%%%%%%%%%%%%%%%%%%%%%%%%%%%%%%%%%%%%%%%%%%%%%%%%%%%%%%%%%%%%%%%%%%%
\subsubsection{Example 1: neutrino masses from the Weinberg operator}
%%%%%%%%%%%%%%%%%%%%%%%%%%%%%%%%%%%%%%%%%%%%%%%%%%%%%%%%%%%%%%%%%%%%%%%%%%%%%%%%%%%%%%%%%%%
We consider an example where the neutrino masses and mixing angles originate directly from the Weinberg operator.
In this model we have no contribution to the mixing from the charged lepton sector. This can be achieved by 
choosing the field content of table 3. In particular we need a flavon $\varphi_T$ transforming as a triplet of $\Gamma_3$ and developing a VEV of the type:
\be
\langle\varphi_T\rangle=(u,0,0)~~~.
\label{vevt}
\ee
Such a VEV breaks $\Gamma_3$ down to the $Z_3$ subgroup generated by $T$.
\begin{table}[h!] 
\centering
\begin{tabular}{|c|c|c|c|c|c|c||c|}
\hline
&$E_1^c$& $E_2^c$& $E_3^c$& $L$& $H_d$& $H_u$&$\varphi_T$\rule[-2ex]{0pt}{5ex}\\
\hline
$SU(2)_L\times U(1)_Y$&$(1,+1)$& $(1,+1)$& $(1,+1)$& $(2,-1/2)$& $(2,-1/2)$& $(2,+1/2)$& $(1,0)$ \rule[-2ex]{0pt}{5ex}\\
\hline
$\Gamma_3\equiv A_4$&$1$& $1''$& $1'$& $3$& $1$& $1$& $3$ \rule[-2ex]{0pt}{5ex}\\
\hline
$k_I$&$k_{E_1}$& $k_{E_2}$& $k_{E_3}$& $k_L$& $k_d$& $k_u$& $k_\varphi$ \rule[-2ex]{0pt}{5ex}\\
\hline
\end{tabular}
\caption{Chiral supermultiplets, transformation properties and weights.}
\label{tabmod3w}
\end{table}
We choose the weights $k_{E_i}$, $k_L$, $k_d$ and $k_\varphi$ such that 
$k_{E_i}+k_L+k_d+k_\varphi=0$. Moreover, to forbid a dependence of the charged lepton masses on $Y(\tau)$ (and a dependence of the Weinberg operator on $\varphi_T$), we take, for instance, $k_\varphi=-3$.
The superpotential for the charged lepton sector reads:
\be
\label{we}
w_e=\alpha~ E_1^c H_d (L~\varphi_T)_1+\beta~ E_2^c H_d (L~\varphi_T)_{1'}+\gamma~E_3^c H_d (L~\varphi_T)_{1''}~~~.
\ee
The VEV of eq. (\ref{vevt}) leads to a diagonal mass matrix for the charged leptons:
\be
\label{mediag}
m_e={\tt diag}(\alpha,\beta,\gamma)u~ v_d~~~.
\ee
The charged lepton masses can be reproduced
by adjusting the parameters $\alpha$, $\beta$ and $\gamma$, with an ambiguity related to the freedom of permuting the eigenvalues.
As a result, the lepton mixing matrix $U_{PMNS}$ is determined up to a permutation of the rows.
Finally, by choosing $k_L=+1$ and $k_u=0$, we uniquely determine the form of the Weinberg operator: 
\be
w_\nu=\dd\frac{1}{\Lambda}(H_u H_u~ L L~ Y)_1
\label{su0}
\ee
\vspace{0.5 cm}
\begin{table}[h!] 
\centering
\begin{tabular}{|c|c|c|c|c|c|c|c|}
\hline
&$r=\frac{\Delta m^2_{sol}}{|\Delta m^2_{atm}|}$ & $\sin^2\theta_{12}$ & $\sin^2\theta_{13}$ & $\sin^2\theta_{23}$ & $\frac{\delta_{CP}}{\pi}$ & $\frac{m_e}{m_\mu}$ & $\frac{m_\mu}{m_\tau}$ \rule[-2ex]{0pt}{5ex}\\
\hline
{\tt best value}& $0.0292$ & $0.297$ & $0.0215$ & $0.5$ & $1.4$ & $0.0048$ & $0.0565$\rule[-2ex]{0pt}{5ex}\\
\hline
{\tt 1$\sigma$ error} & $0.0008$ & $0.017$ & $0.0007$ & $0.1$ & $0.2$ & $0.0002$ & $0.0045$\rule[-2ex]{0pt}{5ex}\\
\hline
\end{tabular}
\caption{Values of observables and their 1$\sigma$ errors used to optimize the model parameters, through a $\chi^2$ scan. Oscillation parameters are from ref. \cite{Capozzi:2017ipn} and ratios of charged lepton masses from ref. \cite{Ross:2007az}. We use $|\Delta m^2_{atm}|=|m_3^2-(m_1^2+m_2^2)/2|$ where $m_i$ are the neutrino masses. The ratios $\frac{m_e}{m_\mu}$ and $\frac{m_\mu}{m_\tau}$ are evaluated at the scale $2\times 10^{16}$ GeV. For $\frac{m_\mu}{m_\tau}$ the average between the values obtained with $\tan\beta=10$ and $\tan\beta=38$ has been used. There is a sizable difference between the allowed 1$\sigma$ ranges of $\sin^2\theta_{23}$ for the cases of normal and inverted ordering.
For simplicity we have adapted the ranges quoted in ref. \cite{Capozzi:2017ipn} and we use a unique range for the two cases. The value of $\frac{\delta_{CP}}{\pi}$ has not been used in the scan.}
\label{data}
\end{table}

\noindent
The superpotential $w=w_e+w_\nu$ depends on the four parameters $\alpha,\beta,\gamma,\Lambda$. 
The charged lepton masses $m_e$, $m_\mu$ and $m_\tau$ are in a one-to-one correspondence with $\alpha$, $\beta$ and $\gamma$, which
can be taken real without loosing generality. The neutrino mass matrix is given by:
\be
m_\nu=
 \left(
 \begin{array}{ccc}
2 Y_1&-Y_3&-Y_2\\
-Y_3&2 Y_2&-Y_1\\
-Y_2&- Y_1& 2  Y_3 
 \end{array}
 \right)\frac{v_u^2}{\Lambda}
\ee
We see that the fourth parameter, $\Lambda$, controls the absolute scale of neutrino masses. 
A remarkable feature of this model is that neutrino mass ratios, lepton mixing angles, Dirac and Majorana phases are completely determined by 
the modulus $\tau$.
We have eight dimensionless physical quantities that do not depend on any coupling constant. 
Assuming the VEV of eq. (\ref{vevt}) for the flavon $\varphi_T$,
they all uniquely depend on $\tau$.
Four of these parameters,  $r=\Delta m^2_{sol}/|\Delta m^2_{atm}|$, $\sin^2\theta_{12}$, $\sin^2\theta_{13}$ and $\sin^2\theta_{23}$, have been measured with good precision. A fifth one, $\delta_{CP}$, starts to be constrained by the present data, see table 4.
It is a significant challenge to reproduce all of them by varying a single complex parameter.
By scanning a portion of the upper complex plane where $\tau$ varies, we 
found that the agreement between predictions and data is optimized by the choice:
\be
\tau=0.0111+0.9946 i~~~,
\ee
giving rise to an inverted neutrino mass ordering. Assuming that the charged lepton sector induces a permutation between the second and the third rows,
the neutrino mass/mixing parameters are predicted to be \footnote{Dirac and Majorana phases are in the PDG convention.}
\be
\label{res312}
\begin{array}{lll}
\dd\frac{\Delta m^2_{sol}}{|\Delta m^2_{atm}|}=0.0292~~~&&\\[0.4 cm]
\sin^2\theta_{12}=0.295~~~&\sin^2\theta_{13}=0.0447~~~&\sin^2\theta_{23}=0.651~~~\\[0.2 cm]
\dd\frac{\delta_{CP}}{\pi}=1.55~~~&\dd\frac{\alpha_{21}}{\pi}=0.22~~~&\dd\frac{\alpha_{31}}{\pi}=1.80~~~. 
\end{array}
\ee
Also the absolute scale of neutrino masses is determined, since $v_u^2/\Lambda$ can be fixed by requiring that the individual square mass differences $|\Delta m^2_{atm}|$ and $\Delta m^2_{sol}$ are reproduced.
We find the central values:
\be
m_1=4.998\times 10^{-2}~eV~~~~~~~~~~m_2=5.071\times 10^{-2}~eV~~~~~~~~~~m_3=7.338\times 10^{-4}~eV~~~.
\ee
Some comments are in order.
\begin{itemize}
\item[$\bullet$]
The value of $\tau$ that minimises the $\chi^2$ is intriguingly close to the self-dual point $\tau=i$ where $S$ is unbroken. It is easy to verify that the self-dual point leads to an inverted-ordering spectrum with $\Delta m^2_{sol}=0$
and relative Majorana phase $\pi$ between the first two neutrino levels. For $\tau=i$ the CP symmetry is unbroken and the non-trivial phases in eq. (\ref{res312}) are entirely generated by the departure of $\tau$ from $i$
\footnote{When $\tau=i$ the mixing matrix $U$ is real and, from eq. (\ref{buono}), we find $\sin^2\theta_{12}=(11-6\sqrt{3})/26\approx 0.02$,  $\sin^2\theta_{23}=(16-4\sqrt{3})/26\approx0.35$ and $|U_{e3}|=(3-\sqrt{3})/6\approx 0.21$.}. 
\item[$\bullet$]
Three observables, $r=\Delta m^2_{sol}/|\Delta m^2_{atm}|$, $\sin^2\theta_{12}$ and $\delta_{CP}$ are within the 1$\sigma$ experimental range. A fourth one, $\sin^2\theta_{23}$, is very close
to the 3$\sigma$ range \cite{Capozzi:2017ipn} for the inverted ordering.
\item[$\bullet$]
The minimum $\chi^2$ is completely dominated by $\sin^2\theta_{13}$, which is many $\sigma$s away from the allowed range. Nevertheless $\theta_{13}$ is predicted to be the smallest angle, in qualitative agreement with the data.
\item[$\bullet$]
The observed value of $\delta_{CP}$ was not included in the data set used to minimise the $\chi^2$, but it turns out to be in very good agreement with the current determination. 
\item[$\bullet$]
The two Majorana phases are predicted and the parameter $|m_{ee}|$ of neutrinoless double beta decay is completely determined.
\item[$\bullet$]
Had we assumed no permutation between the second and third row of the mixing matrix, we would have obtained $\sin^2\theta_{23}=0.349$ and $\delta/\pi=-1.55$, with no change for the other predictions.
\item[$\bullet$]
Notice that modular symmetry and supersymmetry completely determine the superpotential in the lepton sector in terms of the free parameters $\alpha$, $\beta$, $\gamma$ and $\Lambda$.
\end{itemize}
There are no extra contribution we can add to $w$ that cannot be absorbed in a redefinitions of those parameters. In particular the couplings of the modulus $\tau$ to the matter supermultiplets
are fixed to any order in the $\tau$ power expansion and all higher-dimensional operators involving $\tau$ are known as a function
of a finite set of parameters. This remarkable feature has no counterpart in the known models based on discrete symmetries where, in general, to fix all higher-dimensional operators
involving the relevant flavon fields, an infinite number of parameters is needed.
As a consequence, corrections to the above predictions can only come from few sources. One is the vacuum alignment. In this example we have kept fixed the VEV of $\varphi_T$. Additional, small contributions to the mixing can be expected
if we relax this assumption.  Other sources of corrections can be either supersymmetry breaking contributions or corrections to the Kahler potential.
A candidate modification of the Kahler potential (\ref{kalex}) able to induce corrections to the above results  is an additive contribution depending explicitly on both the matter supermultiplets and on the modular forms $Y_i(\tau)$.  
The problem of accounting for these additional terms will be analyzed elsewhere.
The properties discussed here are not specific of the example under examination, but rather general features of the setup proposed here.
A small correction to the model, coming for instance from the charged lepton sector, might bring the model to agreement with $\theta_{13}$, perhaps without spoiling
the predictions for the remaining observables. A more quantitative analysis will be pursued elsewhere. We consider the above zeroth-order approximation as an excellent starting point.
%%%%%%%%%%%%%%%%%%%%%%%%%%%%%%%%%%%%%%%%%%%%%%%%%%%%%%%%%%%%%%%%%%%%%%%%%%%%%%%%%%%%%%%%%%%
\subsubsection{Example 2: neutrino masses from the see-saw mechanism}
%%%%%%%%%%%%%%%%%%%%%%%%%%%%%%%%%%%%%%%%%%%%%%%%%%%%%%%%%%%%%%%%%%%%%%%%%%%%%%%%%%%%%%%%%%%
In this example neutrino masses and mixing angles come from the see-saw mechanism. The field content is the one of table 5 and includes a triplet of chiral supermultiplets $N^c$ describing right-handed neutrinos. 
The charged lepton sector is exactly the same as in the previous example. The relevant superpotential $w_e$ is the one of eq. (\ref{we}). 
As before the weights $k_{E_i}$, $k_L$, $k_d$ and $k_\varphi$ satisfy $k_{E_i}+k_L+k_d+k_\varphi=0$ and we choose $k_\varphi=-3$.
We assume that the flavon $\varphi_T$, transforming as a triplet of $\Gamma_3$, develops the VEV in eq. (\ref{vevt}), which guarantees a diagonal mass matrix for the charged leptons, as in eq. (\ref{mediag}).
There is no contribution to the mixing from $w_e$ and the charged lepton masses are in a one-to-one correspondence with the parameters $\alpha$, $\beta$ and $\gamma$. Also in this case there is an ambiguity related to the freedom of permuting the eigenvalues of $m_e$
and the lepton mixing matrix $U_{PMNS}$ is determined up to a permutation of the rows.
\begin{table}[h!] 
\centering
\begin{tabular}{|c|c|c|c|c|c||c|}
\hline
&$(E_1^c,E_2^c,E_3^c)$&$N^c$& $L$& $H_d$& $H_u$&$\varphi_T$\rule[-2ex]{0pt}{5ex}\\
\hline
$SU(2)_L\times U(1)_Y$&$(1,+1)$ & $(1,0)$ & $(2,-1/2)$& $(2,-1/2)$& $(2,+1/2)$& $(1,0)$ \rule[-2ex]{0pt}{5ex}\\
\hline
$\Gamma_3\equiv A_4$&$(1,1'',1')$& $3$& $3$& $1$& $1$& $3$ \rule[-2ex]{0pt}{5ex}\\
\hline
$k_I$&$(k_{E_1},k_{E_2},k_{E_3})$& $k_N$& $k_L$& $k_d$& $k_u$& $k_\varphi$ \rule[-2ex]{0pt}{5ex}\\
\hline
\end{tabular}
\caption{Chiral supermultiplets, transformation properties and weights.}
\label{tabmod5w}
\end{table}
By choosing the weights $k_N+k_u+k_L=0$ and $k_N=+1$, the superpotential of the neutrino sector is given by:
\be
w_\nu=g~(N^c H_u L)_1+\Lambda (N^c N^c Y)_1
\ee
With the above choice of weights this the only gauge-invariant holomorphic polynomial singlet under $\Gamma_3$ and satisfying the constraint of eq. (\ref{compensate}).
Using a vector notation $({E^c}^T=(E_1^c,E_2^c,E_3^c),...)$, we can write the superpotential $w=w_e+w_\nu$ as:
\be
\begin{array}{l}
w={E^c}^T {\cal Y}_e H_d L+{N^c}^T {\cal Y}_\nu H_u L+ {N^c}^T {\cal M}_R N^c \\[0.6 cm]
 {\cal Y}_e=
 \left(
 \begin{array}{ccc}
\alpha~ \varphi_{T1}&\alpha~ \varphi_{T3}&\alpha~ \varphi_{T2}\\
\beta~ \varphi_{T2}& \beta~ \varphi_{T1}&\beta~ \varphi_{T3}\\
\gamma~ \varphi_{T3}&\gamma~ \varphi_{T2}&\gamma~ \varphi_{T1}
 \end{array}
 \right)\\[0.8 cm]
 {\cal Y}_\nu=g
 \left(
 \begin{array}{ccc}
1&0&0\\
0&0&1\\
0&1&0
 \end{array}
 \right)\\[0.8 cm]
{\cal  M}_R=
 \left(
 \begin{array}{ccc}
2 Y_1&-Y_3&-Y_2\\
-Y_3&2 Y_2&-Y_1\\
-Y_2&- Y_1& 2  Y_3 
 \end{array}
 \right)\Lambda
\end{array}
\ee
The light neutrino mass matrix $m_\nu$ derived from the see-saw and the charged lepton mass matrix $m_e$ read
\be
\begin{array}{l}
m_\nu= - {\cal Y}_\nu^T {\cal  M}_R^{-1} {\cal Y}_\nu \langle H_u\rangle^2\\[0.2 cm]
m_e= {\cal Y}_e \langle H_d\rangle~~~.
\end{array}
\ee
The number of independent low-energy parameters is the same as in the previous example, since $g$ only enters in the combination $g^2/\Lambda$ \footnote{The phases of $g$ and $\Lambda$ are unobservable.}.
A good agreement between data and predictions is obtained with the choice:
\be
\tau=-0.195 + 1.0636 i
\ee
giving rise to a normal neutrino mass ordering. Assuming that the charged lepton sector induces a permutation between the second and the third rows,
the neutrino mass/mixing parameters are predicted to be:
\be
\label{res313}
\begin{array}{lll}
\dd\frac{\Delta m^2_{sol}}{|\Delta m^2_{atm}|}=0.0280~~~&&\\[0.4 cm]
\sin^2\theta_{12}=0.291~~~&\sin^2\theta_{13}=0.0486~~~&\sin^2\theta_{23}=0.331~~~\\[0.2 cm]
\dd\frac{\delta_{CP}}{\pi}=1.47~~~&\dd\frac{\alpha_{21}}{\pi}=1.83~~~&\dd\frac{\alpha_{31}}{\pi}=1.26~~~. 
\end{array}
\ee
Also the absolute scale of neutrino masses is determined, since $g^2 v_u^2/\Lambda$ can be fixed by requiring that the individual square mass differences $|\Delta m^2_{atm}|$ and $\Delta m^2_{sol}$ are reproduced.
We find the central values:
\be
m_1=1.096\times 10^{-2}~eV~~~~~~~~~~m_2=1.387\times 10^{-2}~eV~~~~~~~~~~m_3=5.231\times 10^{-2}~eV~~~.
\ee
Most of the comments on the previous model apply to the present example as well. The distinctive feature of the model is the normal ordering of neutrino masses. Beyond the prediction of $\theta_{13}$, which deviates by many standard deviations from the experimental range, also the value of $\sin^2\theta_{23}$ is slightly below the 3$\sigma$ allowed range for normal ordering. 
Nevertheless $\theta_{13}$ is the smallest angle and $\theta_{23}$ the largest one, in qualitative agreement with the data. Corrections coming either from the Kahler potential or from the
VEV of the flavon $\varphi_T$ might restore the full agreement between data and predictions.
%%%%%%%%%%%%%%%%%%%%%%%%%%%%%%%%%%%%%%%%%%%%%%%%%%%%%%%%%%%%%%%%%%%%%%%%%%%%%%%%%%%%%%%%%%%
\subsubsection{Example 3: no flavons}
%%%%%%%%%%%%%%%%%%%%%%%%%%%%%%%%%%%%%%%%%%%%%%%%%%%%%%%%%%%%%%%%%%%%%%%%%%%%%%%%%%%%%%%%%%%
The following example can be regarded as minimal, in that it makes no use of any flavon field other than the modulus $\tau$.
We chose for lepton and Higgs supermultiplets the transformations of table \ref{tabmod3}.
 \begin{table}[b!] 
\centering
\begin{tabular}{|c|c|c|c|c|c|c|c|}
\hline
&$E_1^c$& $E_2^c$& $E_3^c$& $N^c$& $L$& $H_d$& $H_u$\rule[-2ex]{0pt}{5ex}\\
\hline
$SU(2)_L\times U(1)_Y$&$(1,+1)$& $(1,+1)$& $(1,+1)$& $(1,0)$& $(2,-1/2)$& $(2,-1/2)$& $(2,+1/2)$\rule[-2ex]{0pt}{5ex}\\
\hline
$\Gamma_3\equiv A_4$&$1$& $1''$& $1'$& $3$& $3$& $1$& $1$\rule[-2ex]{0pt}{5ex}\\
\hline
$k_I$&$k_{E_1}$& $k_{E_2}$& $k_{E_3}$& $k_N$& $k_L$& $k_d$& $k_u$ \rule[-2ex]{0pt}{5ex}\\
\hline
\end{tabular}
\caption{Chiral supermultiplets, transformation properties and weights.}
\label{tabmod3}
\end{table}
We can exploit the fact that the whole ring ${\cal M}(\Gamma(3))$ is generated by the $A_4$ triplet $Y_i(\tau)$ $(i=1,2,3)$, to cast the general superpotential of eq. (\ref{su1}) in the form:
\be
\begin{array}{l}
w=w_e+w_\nu\\[0.2 cm]
w_e=\alpha~ E_1^c H_d (L~ Y^{a_1})_1+\beta~ E_2^c H_d (L~ Y^{a_2})_{1'}+\gamma~E_3^c H_d (L~ Y^{a_3})_{1''}~~~,\\[0.2 cm]
w_\nu=g(N^c H_u L~ Y^b)_1+\Lambda (N^c N^c Y^c)_1
\end{array}
\label{su}
\ee
where $Y^a$ denotes $a$ insertions of the basic modular forms $Y_i(\tau)$ and $(...)_r$ $(r=1,1',1'',3)$ stands for the $r$ representation of $A_4$. It can occur that there are several independent combinations of the type $(...)_r$. In this case a sum over all contributions with arbitrary coefficients is understood.
The invariance of $w$ under modular transformations implies the following relations for the weights:
\be
\left\{
\begin{array}{l}
2 a_i=k_{E_i}+k_d+k_L~~~~~(i=1,2,3)\\
2 b=k_N+k_u+k_L\\
2c=2 k_N\\
\end{array}
\right.
\ee
To get canonical kinetic terms we have to rescale the supermultiplets of the theory. The effect of such rescaling can be absorbed into the arbitrary coefficients $\alpha$, $\beta$, $\gamma$, $g$ and $\Lambda$.
We now proceed by selecting one example with specific choices of the weights $k_I$.
By choosing $k_{u,d}=0$ and $k_{E_i}=k_N=k_L=1$, we have $a_i=b=c=1$ and all couplings are linear in $Y_i$:
\be
\begin{array}{l}
w_e=\alpha~ E_1^c H_d (L~ Y)_1+\beta~ E_2^c H_d (L~ Y)_{1'}+\gamma~E_3^c H_d (L~ Y)_{1''}~~~,\\[0.2 cm]
w_\nu=g(N^c H_u L~ Y)_1+\Lambda (N^c N^c Y)_1
\end{array}
\ee
Using a vector notation $({E^c}^T=(E_1^c,E_2^c,E_3^c),...)$, we can write:
\be
\begin{array}{l}
w={E^c}^T {\cal Y}_e H_d L+{N^c}^T {\cal Y}_\nu H_u L+ {N^c}^T {\cal M}_R N^c \\[0.6 cm]
 {\cal Y}_e=
 \left(
 \begin{array}{ccc}
\alpha Y_1&\alpha Y_3&\alpha Y_2\\
\beta Y_2& \beta Y_1&\beta Y_3\\
\gamma Y_3&\gamma Y_2&\gamma Y_1 
 \end{array}
 \right)\\[0.8 cm]
 {\cal Y}_\nu=
 \left(
 \begin{array}{ccc}
2 g_1 Y_1&(-g_1+g_2)Y_3&(-g_1-g_2) Y_2\\
(-g_1-g_2)Y_3& 2 g_1 Y_2&(-g_1+g_2) Y_1\\
(-g_1+g_2)Y_2&(-g_1-g_2) Y_1& 2 g_1 Y_3 
 \end{array}
 \right)\\[0.8 cm]
{\cal  M}_R=
 \left(
 \begin{array}{ccc}
2 Y_1&-Y_3&-Y_2\\
-Y_3&2 Y_2&-Y_1\\
-Y_2&- Y_1& 2  Y_3 
 \end{array}
 \right)\Lambda
\end{array}
\ee
Notice that the contribution $g(N^c H_u L~ Y)_1$ depends on two parameters $g_1$ and $g_2$ corresponding to the two independent invariant singlets that we can form out of $N^c$, $L$ and $Y$.
The light neutrino mass matrix $m_\nu$ derived from the see-saw and the charged lepton mass matrix $m_e$ read
\be
\begin{array}{l}
m_\nu= - {\cal Y}_\nu^T {\cal  M}_R^{-1} {\cal Y}_\nu \langle H_u\rangle^2\\[0.2 cm]
m_e= {\cal Y}_e \langle H_d\rangle~~~.
\end{array}
\ee
From the matrices $m_e$ and $m_\nu$  we can compute lepton masses, mixing angles and phases in terms of the effective parameters: $\alpha$, $\beta$, $\gamma$, $g_{1,2}$ and $\tau$.
We focus on mass ratios and,  without loss of generality, we can set the spurion $\Lambda$ to one. Moreover, by exploiting field redefinitions, $\alpha$, $\beta$, $\gamma$ and $g_1$ can be taken real.
Notice that the light neutrino mass matrix $m_\nu$ depends on a single complex parameter $g_2/g_1$, up to an overall factor that does not affect mass ratios and mixing angles.
Mass ratios among charged leptons depend on $\alpha/\gamma$ and $\beta/\gamma$.
By scanning the parameter space of the model we could identify the point $\alpha/\gamma=15.3$, $\beta/\gamma=0.054$, $g_2/g_1=0.029 i$, $\tau=0.008+ 0.98 i$ which
better approaches the data, collected in table 4, though not providing a realistic set of observables. We find that neutrino masses have inverted ordering. Mass ratios are in good agreement with the data:
\be
\frac{\Delta m^2_{sol}}{|\Delta m^2_{atm}|}=0.0292~~~,~~~~~~~\frac{m_e}{m_\mu}=0.0048~~~,~~~~~~~\frac{m_\mu}{m_\tau}=0.0567~~~.
\ee
The mixing angles are:
\be
\sin^2\theta_{12}=0.459~~~,~~~~~~~\sin^2\theta_{13}=0.001~~~,~~~~~~~\sin^2\theta_{23}=0.749~~~.
\ee
An appealing feature of this pattern is that it predicts large solar and atmospheric mixing angles and a small reactor angle in qualitative agreement with the observations.
Nevertheless, given the present experimental accuracy this set of angles is clearly excluded by many standard deviations. For completeness
we also list the Dirac and Majorana phases in the PDG convention:
\be
\frac{\delta_{CP}}{\pi}=1.25~~~,~~~~~~~\frac{\alpha_{21}}{\pi}=1.04~~~,~~~~~~~\frac{\alpha_{31}}{\pi}=1.02~~~. 
\ee
The observed value of $\delta_{CP}$ was not included in the data used to search for the best region of the parameter space and it turns out to be in good agreement with the
current determination. As in the previous examples, modular symmetry and supersymmetry completely determine the superpotential in the lepton sector in terms of the free parameters $\alpha$, $\beta$, $\gamma$, $g_{1,2}$ and $\Lambda$.
In this case the model does not include flavons and corrections to the above predictions can only come from two sources. Either from supersymmetry breaking contributions or from corrections to the Kahler potential.
%%%%%%%%%%%%%%%%%%%%%%%%%%%%%%%%%%%%%%%%%%%%%%%%%%%%%%%%%%%%%%%%%%%%%%%%%%%%%%%%%%%%%%%%%%%
\subsection{Symmetry Realizations}
%%%%%%%%%%%%%%%%%%%%%%%%%%%%%%%%%%%%%%%%%%%%%%%%%%%%%%%%%%%%%%%%%%%%%%%%%%%%%%%%%%%%%%%%%%%
Consider the minimal theory defined by eqs. (\ref{action}), (\ref{kalex}) and (\ref{psex}) in the absence of flavon multiplets $\varphi$. It is reminiscent of a non-linear sigma model and 
it is interesting to better understand how the underlying symmetries are effectively realized. If we set the superpotential $w(\Phi)$ to zero, the action becomes invariant under the full
$SL(2,R)$ continuous group. Indeed the action
\be
{\cal S}=\int d^4x d^2\theta d^2\bar\theta~ K(\Phi,\bar \Phi)
\ee
where
\be
K(\Phi,\bar \Phi)=-h \log(-i\tau+i\bar\tau)+ \sum_I (-i\tau+i\bar\tau)^{-k_I} |\varphi^{(I)}|^2~~~,
\label{kalex1}
\ee
%\be
%\frac{h}{\langle-i\tau+i\bar\tau\rangle^2}\partial_\mu \bar\tau\partial^\mu \tau+\sum_I \frac{\partial_\mu  \overline{\varphi}^{(I)}\partial^\mu \varphi^{(I)}}{\langle-i\tau+i\bar\tau\rangle^{k_I}}~~~~~~~~~~({\tt Im}\tau>0)
%\ee
is invariant under $SL(2,R)$, with no restriction on the real parameters $a$, $b$, $c$, $d$ other than $ad-bc=1$. Starting from a generic $\tau$ $({\tt Im}\tau>0)$, it is always possible
to reach $i$ by a transformation of $SL(2,R)$ and we can choose $\tau=i$ as representative of the vacuum configuration.
Such configuration has an $SO(2)$ invariance, since
\be
\frac{\cos\alpha~ i+\sin\alpha}{-\sin\alpha~ i+ \cos\alpha}=i~~~.
\ee
Hence the modulus $\tau$ parametrizes the coset space $SL(2,R)/SO(2)$.
In this limit $SL(2,R)$ is non-linearly realized and the theory describes the spontaneous breaking
of $SL(2,R)$ down to the subgroup $SO(2)$, $\tau$ describing the corresponding massless Goldstone bosons. 

In the full theory, which includes a non-vanishing superpotential, the continuous symmetry $SL(2,R)$ is explicitly broken down to $SL(2,Z)$.
Here the group $\Gamma_N$ comes into play. Level-$N$ modular forms and matter supermultiplets transform in representations of the groups $\Gamma_N$.
Modular forms depend on the modulus $\tau$ and, in the examples considered so far, the symmetry $\Gamma_N$ is always in the broken phase, as in the case of non-linearly realised continuous symmetries.
The analogy can be pushed further. In the continuous case non-linear realizations can be defined by imposing invariant constraints
on the field space of the theory, such as for instance $\varphi^T \varphi=M^2$ $(M^2> 0)$, the multiplet $\varphi$ transforming in the fundamental of $SO(N)$. 
This constraint induces the breaking of $SO(N)$ down to $SO(N-1)$. Consider the case of a supersymmetric modular-invariant theory with matter multiplets in representations of $\Gamma_3$,
as in Section 3.1. We have seen that the three modular forms
of level 3 and weight 2 obey the relation:
\be
(YY)_{1''}=Y_2^2+2 Y_1Y_3=0~~~.
\label{constraint}
\ee
The left-hand side is covariant, since $(YY)_{1''}$ transforms with a phase factor under $A_4$. Thus the constraint itself 
is left invariant by $A_4$ transformations and induces a restriction on the field space similar to that induced by $\varphi^T \varphi=M^2$
in the case of $SO(N)$. 
For generic values of $Y(\tau)$ satisfying the relation (\ref{constraint}),
the symmetry $\Gamma_3$ is completely broken. Only for the special values related to $\tau=i$ or $\tau=i\infty$ by a $\Gamma_3$ transformation,
a subgroup of $\Gamma_3$ is left unbroken.
Therefore, in some sense, $\Gamma_3$ is non-linearly realised in the models considered here.

Of course there are several differences with the continuous case. In the discrete case there are no Goldstone bosons and
the modulus $\tau$ is expected to be massive.  Non-linear covariant constraints of the type (\ref{constraint}) do not define low-energy effective
theories in the discrete case. These theories should rather be viewed as consistent truncations of some ultraviolet completion. 
In general, as in the case of the models under discussions, covariant constraints leaves no non-trivial residual symmetry. 
We can nevertheless identify the cases in which a non-trivial residual symmetry group survives, by discussing the properties of 
the so-called orbit space, the space spanned by the invariants of the group.  We do this in the remaining part of this work.
%%%%%%%%%%%%%%%%%%%%%%%%%%%%%%%%%%%%%%%%%%%%%%%%%%%%%%%%%%%%%%%%%%%%%%%%%%%%%%%%%%%%%%%%%%%
\section{Non-linear realisations of discrete symmetries}
%%%%%%%%%%%%%%%%%%%%%%%%%%%%%%%%%%%%%%%%%%%%%%%%%%%%%%%%%%%%%%%%%%%%%%%%%%%%%%%%%%%%%%%%%%%
Consider a Lagrangian ${\cal L}(\varphi,\psi)$ invariant under the action of a discrete group $G_d$, depending on a set of scalar fields $\varphi$ responsible for the spontaneous breaking of $G_d$,
plus additional fields $\psi$ that do not play a role in the symmetry breaking. As in the case of a continuous symmetry, we can restrict the field space ${\cal M}$ to which $\varphi$ belong
by means of constraints invariant under the action of $G_d$. 
We classify these constraints into two types:
\begin{itemize}
\item[1.]
constraints of the type $I$, $I$ standing for {\em invariant}, are relations such as:
\be
I_i(\varphi)=0~~~~~~~~~~~~~(i=1,...,n)~~~~~~~{\tt [type~I]}
\label{CI}
\ee
where $I_i(\varphi)$ are invariants under $G_d$. 
\item[2.]
constraints of the type $C$, $C$ standing for {\em covariant}, are defined by the conditions:
\be
C_i(\varphi)=0~~~~~~~~~~~~~(i=1,...,m)~~~~~~~{\tt [type~C]}
\ee
where $C_i(\varphi)$ are covariant, but not invariant, combinations of $\varphi$. Here we focus on those 
transforming with a phase factor under $G_d$
\be
C_i(\varphi')=e^{i \alpha_i}C_i(\varphi)~~~,
\ee
where $\alpha_i$ depend on the $G_d$ transformation.
\end{itemize}
We will consider polynomial invariants, that is $I_i(\varphi)$ are polynomials in the components of the multiplet $\varphi$.
The ring of invariant polynomials is infinite, but it is generated by a finite number of invariants $\gamma_\alpha(\varphi)$, which means that any invariant polynomial can be written as a polynomial in $\gamma_\alpha$.
The invariants $\gamma_\alpha$ might be related by a number of algebraic relations, or syzygies, ${\cal Z}_s(\gamma)=0$.  Also in the case of $C$ constraints we will consider only polynomial expressions.
Constraints of type $I$ are those usually considered to define non-linear realisations of continuous symmetries.
In the discrete case there are other possibilities to restrict the field space ${\cal M}$ in an invariant way.
While a (connected, semisimple) Lie group has only the trivial one-dimensional representation, a discrete group can possess a number of non-invariant one-dimensional representations, or singlets, $\chi_i$. 
Therefore the invariant conditions $\chi_i=0$ have no counterpart in (connected, semisimple) Lie groups and give rise of constraints of type $C$. Notice that both conditions of type $I$ and type $C$ are left invariant by 
transformations of $G_d$.

In this Section we would like to analyse the pattern of symmetry breaking induced by the above constraints. To this purpose it is useful to inspect the orbits
of the group, {\em i.e.} the set of points in the field space ${\cal M}$ that are related by group transformations. Each point of a given orbit has the same little group, up to a conjugation.
The union of orbits having isomorphic little groups forms a stratum. The full field space ${\cal M}$ is partitioned into several strata. For instance the origin of ${\cal M}$ is the stratum of type $G_d$, since for $\varphi=0$ the symmetry is unbroken.
Most of the field space ${\cal M}$ is made of orbits having trivial little group, {\em i.e.} the symmetry $G_d$ is completely broken. This subset of ${\cal M}$ is called principal stratum.
To analyse the symmetry breaking pattern induced by a constraint, it is sufficient to determine the orbit type of the strata affected by the constraint.
An even more effective picture is obtained by moving to the orbit space ${\cal M}_I$, spanned by the invariants $\gamma_\alpha$ of the theory. 
A whole orbit of ${\cal M}$ is mapped into a single point of ${\cal M}_I$. The crucial property of ${\cal M}_I$ is that
while ${\cal M}$ has no boundaries, ${\cal M}_I$ has boundaries that describe the possible breaking chains of the group.
The tools that allow to characterise the orbit space ${\cal M}_I$ are the Jacobian matrix \cite{Cabibbo:1970rza}
\be
J\equiv\frac{\partial \gamma}{\partial\varphi}~~~,
\ee
and the so-called ${\cal P}$-matrix \cite{Talamini:2006wd}
\be
{\cal P}=JJ^T~~~.
\ee
The manifold ${\cal M}_I$ is identified by the requirement that the matrix ${\cal P}$ is positive semidefinite, resulting in a set of inequalities involving the invariants $\gamma_\alpha$.
The boundaries of ${\cal M}_I$ can be found by studying the rank of $J$.
In the interior of ${\cal M}_I$ the matrix $J$ has maximum rank $r_{max}$. This region corresponds to the principal stratum where $G_d$ is completely broken. On the boundaries ${\cal P}$ has some vanishing eigenvalue
and the rank of $J$ is reduced.  If the dimension of ${\cal M}_I$ is $d$, in general we have $(d-1)$-dimensional boundaries where ${\tt rank}(J)=r_{max}-1$. They corresponding to strata of type $G'_d$, with  $G'_d\subset G_d$.
These boundaries meet along $(d-2)$-dimensional spaces, where ${\tt rank}(J)=r_{max}-2$. They correspond to strata of type $G''_d$, with $G'_d\subset G''_d\subset G_d$. And so on, until the 1-dimensional boundaries
meet in a point corresponding to the stratum of type $G_d$. Thus the symmetry breaking induced by a constraint can be immediately visualised by identifying
the region of ${\cal M}_I$ involved by the constraint. Here we provide some examples, by discussing the case of the groups $S_3$ and $A_4$.
%%%%%%%%%%%%%%%%%%%%%%%%%%%%%%%%%%%%%%%%%%%%%%%%%%%%%%%%%%%%%%%%%%%%%%%%%%%%%%%%%%%%%%%%%%%
\subsection{Covariant constraints in $S_3$}
%%%%%%%%%%%%%%%%%%%%%%%%%%%%%%%%%%%%%%%%%%%%%%%%%%%%%%%%%%%%%%%%%%%%%%%%%%%%%%%%%%%%%%%%%%%
$S_3$ is the group of permutations of three objects. It is also the symmetry group of an equilateral triangle, generated by a reflection $a$ with respect to an axis crossing a vertex and orthogonal to the opposite side
and a $2\pi/3$ rotation $ab$ around the center. In terms of the two generators $a$ and $b$ its six elements are given by
\be
S_3=\{e,a,b,ab,ba,bab\}~~~.
\ee
The elements $(ab,ba)$ are of order three and the elements $(a,b,bab)$ are of order two.
They fall into three distinct conjugacy classes $C_1=\{e\}$, $C_2=\{ab,ba\}$, $C_3=\{a,b,bab\}$ and the three irreducible representations of the group are an invariant singlet $1$, a singlet $1'$ and a doublet $2$.
A real basis for the generators $a$ and $b$ is given by:
\be
\begin{array}{lll}
1:
&
a=1
&
b=1
\\
1':
&
a=-1
&
b=-1
\\
2:
&
a=\left(
\begin{array}{cc}
1&0\\
0&-1
\end{array}
\right)
&
b=\left(
\begin{array}{cc}
-\frac{1}{2}&-\frac{\sqrt{3}}{2}\\
-\frac{\sqrt{3}}{2}&\frac{1}{2}
\end{array}
\right)~~~.
\end{array}
\ee
Given two doublets $\varphi=(\varphi_1,\varphi_2)$ and $\psi=(\psi_1,\psi_2)$, we have
\bea
(\varphi\psi)_1&=&\varphi_1\psi_1+\varphi_2\psi_2\nonumber\\
(\varphi\psi)_{1'}&=&\varphi_1\psi_2-\varphi_2\psi_1\\
(\varphi\psi)_2&=&
\left(
\begin{array}{c}
-\varphi_1\psi_1+\varphi_2\psi_2\\
\varphi_1\psi_2+\varphi_2\psi_1
\end{array}
\right)~~~.\nonumber
\eea
We consider a theory invariant under $S_3$ and depending on a single real doublet $\varphi(x)$ in the scalar sector. 
The ring of invariant polynomials is generated by the two basic invariants \footnote{The Molien function of $S_3$ in the doublet representation is $M(x)=1/(1-x^2)(1-x^3)$. It follows that there are two independent invariants, one quadratic and one cubic \cite{Ramond}.}:
\be
\gamma_1=(\varphi\varphi)_1~~~,~~~~~~~\gamma_2=(\varphi\varphi\varphi)_1~~~.
\ee
In our real basis these read:
\be
\gamma_1=\varphi_1^2+\varphi_2^2~~~,~~~~~~~\gamma_2=\varphi_1^3-3\varphi_1\varphi_2^2~~~.
\ee
The invariants $\gamma_{1,2}$ are independent: there are no algebraic relations among them.
Examples of possible constraints of type $I$ are
\be
\varphi_1^2+\varphi_2^2=M_1^2~~~,
\label{sc1}
\ee
\be
\varphi_1^3-3\varphi_1\varphi_2^2=M_2^3~~~.
\label{sc11}
\ee
We notice that the constraint (\ref{sc1}) is invariant under the full orthogonal group $O(2)$ and is analogue to the constraint defining a nonlinear sigma model $G/H$, $G=O(2)$ and $H=\{e\}$. In general, imposing these constraints breaks completely the group $S_3$, in the sense that a generic pair $(\varphi_1,\varphi_2)$ satisfying
(\ref{sc1}) or (\ref{sc11}) or both is not left invariant by any subgroup of $S_3$. These constraints might still be useful in the model building, since they limit the possible set of vacuum configuration of the theory. 

The group $S_3$ has a unique one-dimensional non invariant representation, the singlet $1'$. One can verify that the most general combination of $(\varphi_1,\varphi_2)$
transforming as a $1'$ can be written as
\be
I(\varphi)\chi(\varphi)~~~,
\label{f1}
\ee
where $I(\varphi)$ is an $S_3$ invariant while $\chi(\varphi)$ is given by
\be
\chi(\varphi)=(\varphi\varphi\varphi)_{1'}=3 \varphi_1^2\varphi_2-\varphi_2^3~~~,
\ee
where in the last equality we display the explicit expression in the real basis.
Thus the unique non-trivial example of type $C$ constraint for $S_3$ with a real doublet $\varphi$ is
\be
3 \varphi_1^2\varphi_2-\varphi_2^3=0~~~.
\label{constrS3}
\ee
This has no analogue in $O(2)$. It defines a domain ${\cal D}$, contained in the full field space ${\cal M}\equiv R^2$, with the property of being invariant under the action of $S_3$. It is satisfied 
by points belonging to the orbits 
\be
{\cal O}_\xi=\left((\xi,0),(-\xi/2,-\sqrt{3}/2\xi),(-\xi/2,+\sqrt{3}/2\xi)\right)~~~~~~~~~~~~~~~~~-\infty\le \xi\le +\infty~~~.
\label{domain2}
\ee
For $\xi\ne 0$ these orbits have little group isomorphic to $Z_2$ while for $\xi=0$ the little group coincides with $S_3$.
When $\xi$ varies from $-\infty$ to $+\infty$, ${\cal O}_\xi$ describes a domain ${\cal D}$ consisting of three straight lines
that are permuted under the action of $S_3$. The domain ${\cal D}$ is the union of the stratum of type $Z_2$ and the stratum of type $S_3$.
The orbits ${\cal O}_\xi$ are not generic ones. Generic orbits of $S_3$ consist of six points in the field space ${\cal M}$, see Figure 2. 
In generic orbits the symmetry $S_3$ is completely broken. The full field space ${\cal M}$ is partitioned into three strata: the origin, the stratum of type $Z_2$ and the principal stratum, having trivial little group, see Figure 2.
\begin{figure}[!h]
\label{fs}
\centering
\includegraphics[width=0.47\textwidth]{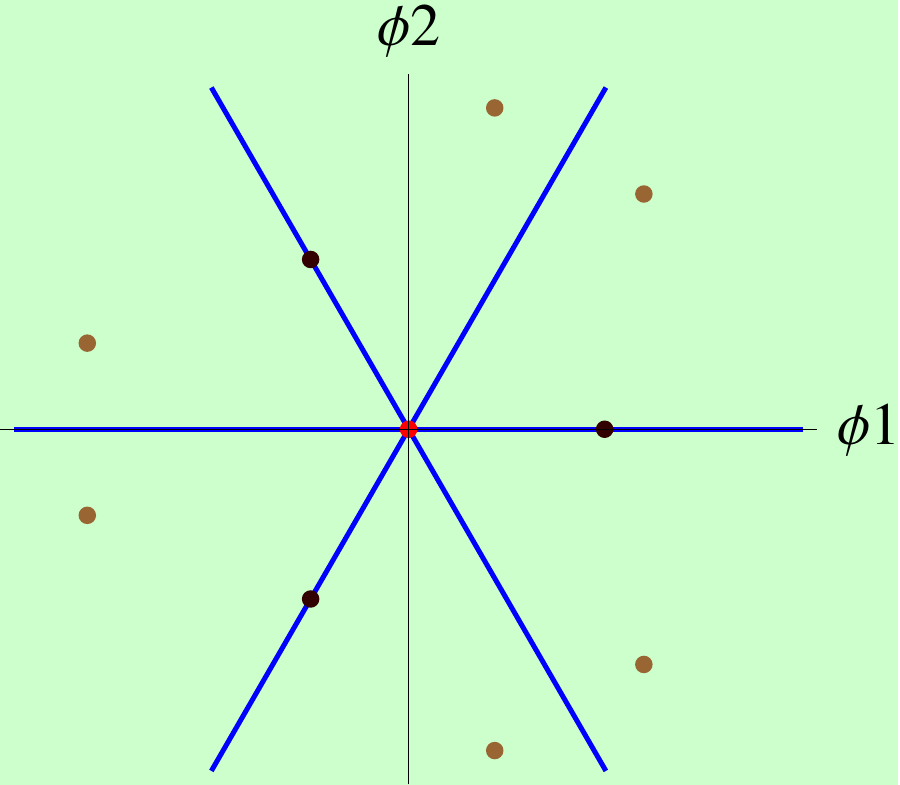}~~~~~
\includegraphics[width=0.42\textwidth]{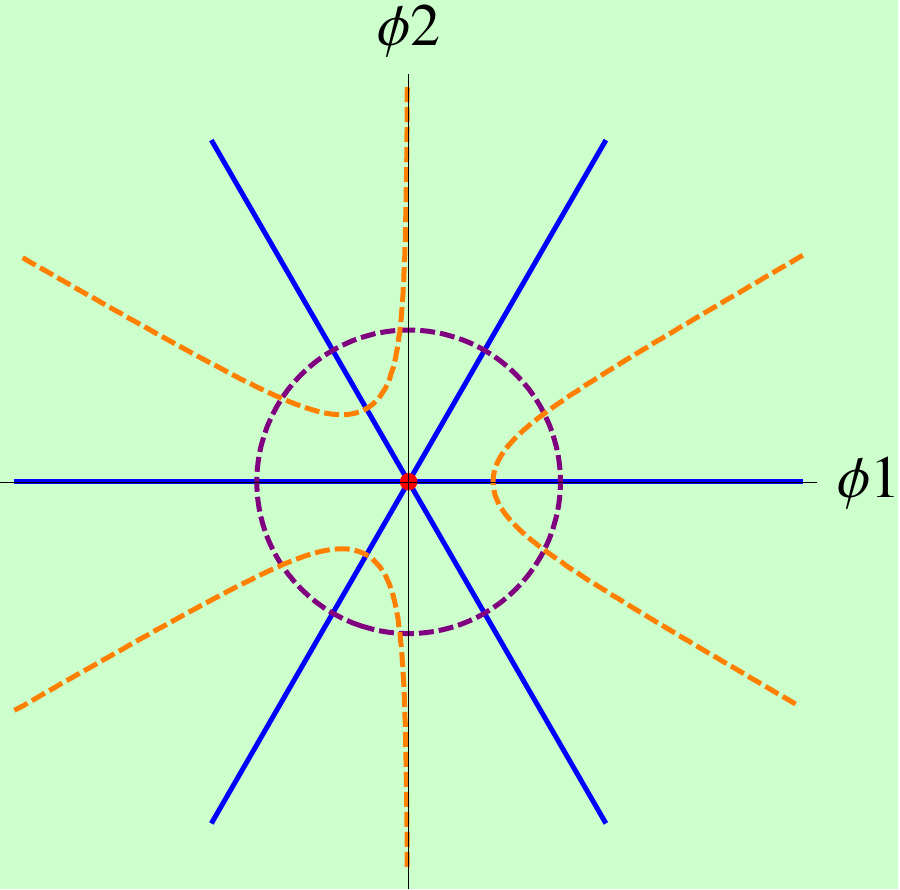}
\caption{{\bf Left plot}: the field space ${\cal M}$, partitioned in strata of different types. The origin (red dot) has little group $S_3$. The set made of the three straight lines is the stratum of type $Z_2$.
For example, an orbit in this stratum is given by the three black points. 
The rest of the plain is the stratum of type $\{ e\}$. This stratum is called principal stratum.  An orbit belonging to this stratum is displayed by the six brown points.
{\bf Right plot}: constraints of type $I$, eq. (\ref{sc1}) (dashed, purple) and eq. (\ref{sc11}) (dashed, orange), represented in the field space ${\cal M}$. In generic points along these lines no $S_3$-subgroup is preserved.}
\end{figure}
Restricting the field space of $\varphi$ through the covariant constraint (\ref{constrS3}) amounts to describe the breaking of $S_3$ down to $Z_2$,
but for the particular case where $\varphi=0$, which leaves $S_3$ unbroken.
The condition (\ref{constrS3}) is invariant under $S_3$ transformations and fields $\varphi=(\varphi_1,\varphi_2)$ obeying such conditions do not abandon the domain ${\cal D}$
when undergoing an $S_3$ transformation. We can say that the discrete symmetry $S_3$ is non-linearly realised, reflecting the fact that the domain ${\cal D}$ cannot be globally parametrized in terms of a single real field variable.
The Lagrangian
\be
{\cal L}=\frac{1}{2}\partial_\mu \varphi_1\partial^\mu \varphi_1+\frac{1}{2}\partial_\mu \varphi_2\partial^\mu \varphi_2-V(\gamma_1,\gamma_2)+\delta {\cal L}(\varphi,\psi)~~~~~~~~~~~~~~\varphi\in {\cal D}
\label{lagr1}
\ee
where $\delta {\cal L}(\varphi,\psi)$ represents $S_3$-invariant terms depending on possible additional fields $\psi$, describes an $S_3$-invariant theory, where $S_3$ is unbroken or spontaneously broken down to $Z_2$.
At variance with non-linear realizations of a continuous group, the scalar field $\xi$ that locally parametrizes the domain ${\cal D}$ is not a Goldstone boson.
Moreover it is expected to have a non-vanishing mass, since it parametrizes a continuous direction linking different group orbits, albeit of the same orbit type. 
In general, we cannot interpret the theory restricted to ${\cal D}$ as a low-energy effective description obtained by integrating out the heavy
degrees of freedom included in $\varphi=(\varphi_1,\varphi_2)$, unless appropriate assumptions on the parameters of the scalar potential $V(\gamma_1,\gamma_2)$ are made.
Thus the theory described by the Lagrangian (\ref{lagr1}) should be viewed as a consistent truncation of the full
theory, rather than a low-energy approximation.

The orbit space ${\cal M}_I$ is spanned by $(\gamma_1,\gamma_2)$. A whole orbit of ${\cal M}$ is mapped into a single point of ${\cal M}_I$, see Figure 3.
The Jacobian matrix $J$ and the ${\cal P}$ matrix read
\be
J\equiv\frac{\partial \gamma}{\partial\varphi}=
\left(
\begin{array}{cc}
2\varphi_1&2\varphi_2\\
3\varphi_1^2-3\varphi_2^2&-6\varphi_1\varphi_2
\end{array}
\right)~~~,~~~~~~~~~~~~~~~~~~~~~~~~~~~~~~
{\cal P}=JJ^T=
\left(
\begin{array}{cc}
4\gamma_1&6\gamma_2\\
6\gamma_2&9\gamma_1^2
\end{array}
\right)~~~.
\ee
The manifold ${\cal M}_I$ is identified by requiring that the matrix ${\cal P}$ is positive semidefinite, which defines the region:
\be
\gamma_1\ge 0~~~,~~~~~~~|\gamma_2|\le \sqrt{\gamma_1^3}~~~.
\ee
We have:
\be
\begin{array}{lcl}
{\tt rank}(J)=1&\leftrightarrow&3\varphi_1^2 \varphi_2-\varphi_2^3=0~~~~(\varphi\ne 0)\\[0.2 cm]
{\tt rank}(J)=0&\leftrightarrow&\varphi= 0
\end{array}
\ee
The 2-dimensional orbit space ${\cal M}_I$ is shown in fig. 3. The conditions $3\varphi_1^2 \varphi_2-\varphi_2^3=0$ $(\varphi\ne 0))$ and ${\tt rank}(J)=1$ are equivalent and identify the 1-dimensional boundary, corresponding to the stratum of type $Z_2$.
The rank of $J$ vanishes for $\varphi=0$, where $\gamma_1=\gamma_2=0$. Thus the origin of  ${\cal M}_I$ represents the stratum of type $S_3$. In the interior of ${\cal M}_I$, the principal  stratum, we have ${\tt rank}(J)=2$ and $S_3$ is completely broken. 

\begin{figure}[h!]
\centering
\includegraphics[width=0.5\textwidth]{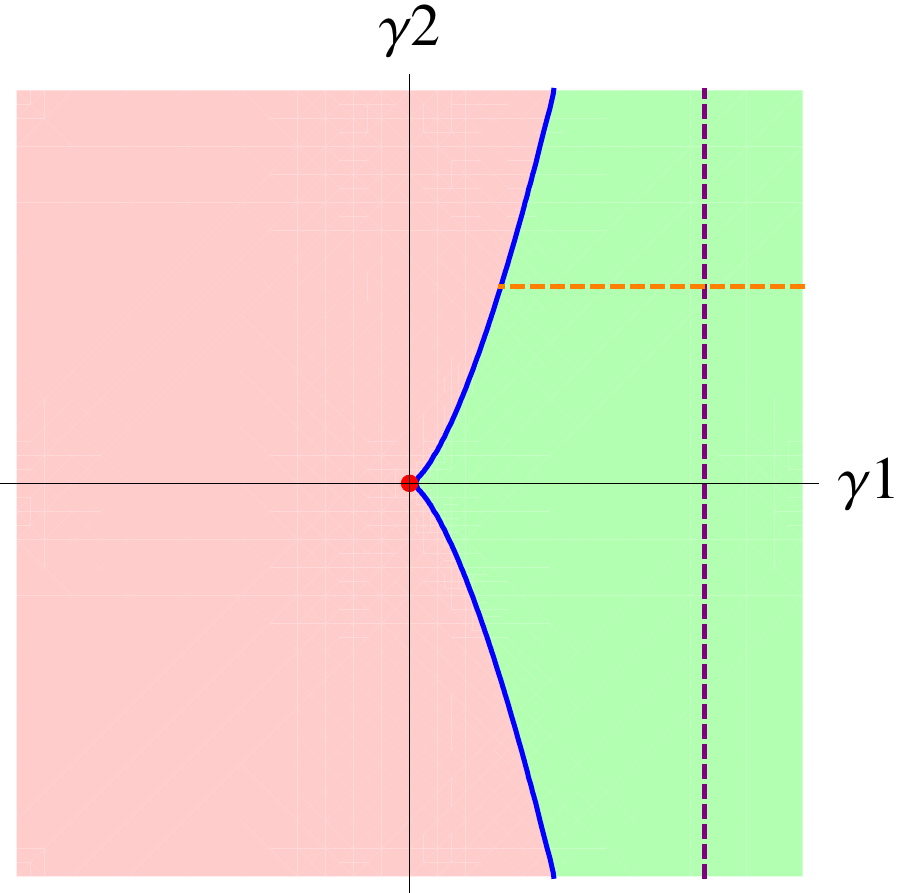}
\caption{The space of invariants ${\cal M}_I$. Each orbit of the field space ${\cal M}$ is mapped into a single point of  ${\cal M}_I$. The origin of  ${\cal M}$
goes into the origin of  ${\cal M}_I$ (red point). The stratum of type $Z_]$, composed of the three straight lines of fig. 2, is mapped into the boundary
of  ${\cal M}_I$ represented by a blue line. The principal stratum is mapped into the region to the right of the blue line. The region to the left of the blue line
is not allowed (see text). The vertical (horizontal) dashed line represents the constraint of eq. (\ref{sc1}) (eq. (\ref{sc11})).}
\end{figure}

Even when we do not make use of constraints to restrict the field space of the theory, 
the above conditions can still be very useful to identify extrema of the scalar potential $V(\gamma_1,\gamma_2)$ in the unrestricted field space ${\cal M}$.
Indeed the boundary of ${\cal M}_I$ provides an important tool to characterize the extrema of $V$ \cite{Cabibbo:1970rza}. The extremum condition reads:
\be
\delta V=\sum_i \frac{\partial V}{\partial \gamma_i}\frac{\partial \gamma_i}{\partial \varphi_\alpha}\delta\varphi_\alpha=\sum_i \frac{\partial V}{\partial \gamma_i}J_{i\alpha}~\delta\varphi_\alpha=0~~~.
\ee
One can prove that extrema of $V$ with respect to the points of the boundary of ${\cal M}_I$ are also extrema of $V$. Moreover extrema on the boundary of ${\cal M}_I$
are more natural than extrema in the interior of ${\cal M}_I$, in the sense that they require the vanishing of a smaller number of derivatives of $V$, 
since on the boundary the rank of the Jacobian matrix $J$ is reduced. Therefore the condition $\chi(\varphi)=3\varphi_1^2 \varphi_2-\varphi_2^3=0$ parametrizes possible natural extrema of the scalar potential. Additional conditions are in general required to guarantee the existence of an extremum on the boundary of ${\cal M}_I$ \cite{Michel:1970mua,Cabibbo:1970rza}. Invariants of continuous groups relevant to particle physics have been extensively used in the literature \cite{Jenkins:2009dy,Degee:2012sk,
Espinosa:2012uu,Alonso:2013nca,Fong:2013dnk}.

The previous example shows a first relation between constraints of type $C$ and the breaking of a discrete symmetry. 
The covariant constraint $3 \varphi_1^2\varphi_2-\varphi_2^3=0$ enforces the breaking of $S_3$ down to a $Z_2$ subgroup.
This is not  a general result, but rather a special outcome related to the fact that with a single real doublet there is only one independent
constraint of type $C$ that can be constructed. To analyse a more general case we move to the group $A_4$.
%%%%%%%%%%%%%%%%%%%%%%%%%%%%%%%%%%%%%%%%%%%%%%%%%%%%%%%%%%%%%%%%%%%%%%%%%%%%%%%%%%%%%%%%%%%
\subsection{Covariant constraints in $A_4$}
%%%%%%%%%%%%%%%%%%%%%%%%%%%%%%%%%%%%%%%%%%%%%%%%%%%%%%%%%%%%%%%%%%%%%%%%%%%%%%%%%%%%%%%%%%%
We replicate the previous analysis in the case of $A_4$, isomorphic to $\Gamma_3$ and of direct interest to modular forms of level 3. The basic properties of $A_4$ are summarised in Appendix C. 
We consider a theory invariant under $A_4$, whose scalar sector depends on a single real triplet $\varphi(x)$. 
As we did in the example of the $S_3$ group, we restrict the field space ${\cal M}=R^3$ by means of constraints. Also in this case we can formulate two types
of constraints. Constraints of type $I$ are of the form $I_i(\varphi)=0$ as in eq. (\ref{CI}), where $I_i(\varphi)$ are polynomial invariants. With a single real triplet the ring of invariant polynomials is generated by the four invariants $\gamma_i$ $(i=1,...,4)$ listed in table \ref{invart}.
These invariants are not algebraically independent. They satisfy the relation, or Syzygy \cite{Ramond}:
\be
{\cal Z}(\gamma)=4\gamma_4^2-2\gamma_3^3+108 \gamma_2^4+\gamma_1^6+36 \gamma_3\gamma_2^2\gamma_1-20 \gamma_2^2\gamma_1^3+5\gamma_3^2\gamma_1^2-4\gamma_3\gamma_1^4=0~~~.
\ee
As we discussed in the case of $S_3$, imposing a generic constraint $I_i(\varphi)=0$ of type $I$ breaks completely the $A_4$ symmetry.
\begin{table}[h] 
\centering
\begin{tabular}{|c|c|c|c|}
\hline
& &{\tt Real basis}&{\tt Complex basis}\\
\hline
$\gamma_1$& $(\varphi^2)_1$& $\varphi_1^2+\varphi_2^2+\varphi_3^2$ & $\varphi_1^2+2\varphi_2\varphi_3$\rule[-2ex]{0pt}{5ex} \\  
\hline
$\gamma_2$& $\frac{1}{6}(\varphi^3)_1$& $\varphi_1\varphi_2\varphi_3$ &$\frac{1}{3\sqrt{3}}\left(\varphi_1^3+\varphi_2^3+\varphi_3^3-3 \varphi_1\varphi_2\varphi_3\right)$ \rule[-3ex]{0pt}{8ex} \\ 
\hline
$\gamma_3$& $\frac{1}{3}\left[(\varphi^2)_1^2+2 (\varphi^2)_{1'}(\varphi^2)_{1''}\right]$& $\varphi_1^4+\varphi_2^4+\varphi_3^4$ &$\frac{1}{3}\left[\varphi_1^4+12 \varphi_1^2\varphi_2\varphi_3\right.$ \rule[-2ex]{0pt}{5ex}\\  
&&&$\left.+4 \varphi_1 (\varphi_2^3+\varphi_3^3) +6 \varphi_2^2\varphi_3^2\right]$\rule[-2ex]{0pt}{5ex}\\
\hline
$\gamma_4$& $-\frac{i}{3\sqrt{3}}\left[(\varphi^2)_{1'}^3-(\varphi^2)_{1''}^3 \right]$ & $(\varphi_1^2-\varphi_2^2)(\varphi_2^2-\varphi_3^2)(\varphi_3^2-\varphi_1^2)$ &
$-\frac{i}{3\sqrt{3}} \left[\varphi_3^6-\varphi_2^6 +6 \varphi_1\varphi_2\varphi_3(\varphi_3^3-\varphi_2^3)\right.$
\rule[-2ex]{0pt}{5ex} \\ 
&&&$\left.-8\varphi_1^3(\varphi_3^3-\varphi_2^3)\right]$\rule[-2ex]{0pt}{5ex}\\
\hline
\end{tabular}
\caption{Invariants that generate the ring of invariant polynomials in $A_4$ and depend on the real triplet $\varphi$. They are not algebraically independent, see the text.}
\label{invart}
\end{table}
We turn to constraints of type $C$, which we formulate by using the non-invariant singlets $\chi_i$ $(i=1,...,4)$ shown in table \ref{nis}.
\begin{table}[h] 
\centering
\begin{tabular}{|c|c|c|c|}
\hline
& &{\tt Real basis}&{\tt Complex basis}\\
\hline
$\chi_1$& $(\varphi^2)_{1'}$& $\varphi_1^2+\omega^2\varphi_2^2+\omega \varphi_3^2$ & $\varphi_3^2+2\varphi_1\varphi_2$\rule[-2ex]{0pt}{5ex} \\  
\hline
$\chi_2$& $(\varphi^2)_{1''}$& $\varphi_1^2+\omega\varphi_2^2+\omega^2 \varphi_3^2$ & $\varphi_2^2+2\varphi_1\varphi_3$\rule[-2ex]{0pt}{5ex} \\  
\hline
$\chi_3$& $(\varphi^4)_{1'}$ & $\varphi_2^2\varphi_3^2+\omega^2 \varphi_3^2\varphi_1^2+\omega \varphi_1^2\varphi_2^2$ &$\frac{1}{3}(\varphi_2^4-2\varphi_2(\varphi_1^3+\varphi_3^3)+3\varphi_1^2\varphi_3^2)$\rule[-2ex]{0pt}{5ex} \\  
\hline
$\chi_4$& $(\varphi^4)_{1''}$ & $\varphi_2^2\varphi_3^2+\omega \varphi_3^2\varphi_1^2+\omega^2 \varphi_1^2\varphi_2^2$ &$\frac{1}{3}(\varphi_3^4-2\varphi_3(\varphi_1^3+\varphi_2^3)+3\varphi_1^2\varphi_2^2)$\rule[-2ex]{0pt}{5ex} \\  
\hline
\end{tabular}
\caption{Some covariant combinations of the real triplet $\varphi$. The product $(\varphi^4)$ contains several non-invariant singlets and $(\varphi^4)_{1',1''}$ represent two possible combinations.}
\label{nis}
\end{table}
By working in the real basis, we start by examining the constraint:
\be
\chi_1=\varphi_1^2+\omega^2\varphi_2^2+\omega \varphi_3^2=0~~~.
\label{c3real}
\ee
By taking the real and the imaginary parts of eq. (\ref{c3real}) we get
\be
\varphi_1^2=\varphi_2^2=\varphi_3^2~~~,
\label{creal}
\ee
which implies  also $\chi_2= 0$.  The constraint is satisfied 
by points belonging to the orbits 
\be
{\cal O}_{2\xi}\equiv\left((\xi,\xi,\xi),(-\xi,\xi,-\xi),(-\xi,-\xi,\xi),(\xi,-\xi,-\xi)\right)~~~~~~~~~~~~~~~~~-\infty\le \xi\le +\infty~~~.
\label{domain2}
\ee
For $\xi\ne 0$ these orbits have little group isomorphic to $Z_3$ while for $\xi=0$ the little group coincides with $A_4$.
When $\xi$ varies from $-\infty$ to $+\infty$, ${\cal O}_{2\xi}$ describes a domain ${\cal D}_2$ consisting of four straight lines
that are permuted under the action of $A_4$. The domain ${\cal D}_2$ is the union of the stratum of type $Z_3$ and the stratum of type $A_4$.
As a whole, the set of four lines is left invariant under $A_4$. 
To complete the partition of the field space ${\cal M}$ into strata, we analyze the the constraint \footnote{The conditions $\chi_3=0$ and $\chi_4=0$ 
have the same solutions, for a real triplet.}
\be
\chi_3=\varphi_2^2\varphi_3^2+\omega^2\varphi_3^2\varphi_1^2+\omega \varphi_1^2\varphi_2^2=0~~~.
\label{chi3real}
\ee
This constraint is solved by all the points in ${\cal D}_2$ and also by points belonging to the orbits
\be
{\cal O}_{4\xi}\equiv\left((\xi,0,0),(-\xi,0,0),(0,\xi,0),(0,-\xi,0),(0,0,\xi),(0,0,-\xi)\right)~~~~~0\le \xi\le +\infty \}~~~.
\label{domainD4}
\ee
For $\xi\ne 0$ these orbits have little group isomorphic to $Z_2$ while for $\xi=0$ the little group coincides with $A_4$.
When $\xi$ varies from $-\infty$ to $+\infty$, ${\cal O}_{4\xi}$ describes a domain ${\cal D}_4$ consisting of three straight lines
that are permuted under the action of $A_4$. Excluding the origin, these lines form a stratum of type $Z_2$. 
Thus the two covariant constraints (\ref{c3real},\ref{chi3real}) provide a partition of the field space ${\cal M}$ into strata. The domains ${\cal D}_2$ and ${\cal D}_4$
describe, for $\xi\ne 0$, strata of type $Z_3$ and $Z_2$ respectively. The origin is the stratum of type $A_4$ and the rest is the principal stratum.

Moving to the space of invariants ${\cal M}_I$, it is easy to check that both the strata ${\cal D}_2$ and ${\cal D}_4$ are mapped into boundaries of ${\cal M}_I$.
Indeed along the orbits ${\cal O}_{2\xi}$ and ${\cal O}_{4\xi}$ the rank of the Jacobian matrix
\be
J^T\equiv\left(\frac{\partial \gamma}{\partial\varphi}\right)^T=
\left(
\begin{array}{cccc}
2\varphi_1&\varphi_2\varphi_3&4\varphi_1^3&2\varphi_1(\varphi_2^2-\varphi_3^2)(\varphi_2^2-2 \varphi_1^2+\varphi_3^2)\\
2\varphi_2&\varphi_3\varphi_1&4\varphi_2^3&2\varphi_2(\varphi_3^2-\varphi_1^2)(\varphi_3^2-2 \varphi_2^2+\varphi_1^2)\\
2\varphi_3&\varphi_1\varphi_2&4\varphi_3^3&2\varphi_3(\varphi_1^2-\varphi_2^2)(\varphi_1^2-2 \varphi_3^2+\varphi_2^2)
\end{array}
\right)~~~,
\ee
is equal to one for $\xi\ne 0$ and zero when $\xi=0$, while for generic orbits the rank of $J$ is equal to three. The space of invariants ${\cal M}_I$ is more complicated
than the one discussed in the $S_3$ case, due to the presence of the Syzygy ${\cal Z}(\gamma)=0$. ${\cal M}_I$ lies on the three-dimensional hyper-surface defined by ${\cal Z}(\gamma)=0$ and it is identified by the requirement that the matrix ${\cal P}=JJ^T$ is positive semidefinite.  There are no two-dimensional boundaries, as it might be expected in a smooth three-dimensional manifold, and ${\cal M_I}$ is a sort of hyperconical surface.

Concerning the applications to fermion masses, the models that we can build using the constraints (\ref{c3real}) and (\ref{chi3real}) are still far from realistic.
In most of the known models specific vacuum alignments among several multiplets of a discrete group are required. For instance, in the lepton sector, typically
we ask distinct residual symmetries in the neutrino and in the charged lepton sectors. If the model contains just a single real triplet of $A_4$,
the above constraints only allow to break $A_4$ down to a $Z_2$ or a $Z_3$ subgroup. 
Nevertheless this approach can be easily generalized to models
with a larger scalar sector and may help exploring different paths in model building. 

An interesting case occurs when a covariant constraint does not correspond to a non-trivial residual symmetry, as it happens in the class of supersymmetric modular-invariant models 
built by using modular forms $Y_i(\tau)$ of level 3. We consider an $A_4$-invariant theory depending on a complex scalar triplet $\varphi=(\varphi_1,\varphi_2,\varphi_3)$ and
for convenience we adopt the complex basis for the three-dimensional representation. We restrict to holomorphic constraints, having in mind the case of supersymmetric theories where $\varphi$ represents
a triplet of chiral supermultiplets of $N=1$ supersymmetry. We can read the relevant invariant and covariant combinations of $\varphi$ in tables \ref{invart} and \ref{nis}, second columns. In particular we focus on the covariant constraint
\be
(\varphi\varphi)_{1''}\equiv \varphi_2^2+2 \varphi_1 \varphi_3=0~~~.
\label{c3}
\ee
This is precisely the algebraic relation met when examining modular forms of level 3 and weight 2.
We see that this constraint does not imply any more $(\varphi\varphi)_{1'}=0$, due to the complex nature of the $\varphi_i$ components. 
Therefore imposing (\ref{c3}) does not necessarily enforce the breaking of $A_4$ into $Z_3$, as in the case of a real triplet. Indeed this is what we found when exploring the models 
of Section 3.1. Generic orbits of the field space satisfying this constraint belong to the principal stratum and are mapped into the interior of the orbit space, where the 
symmetry is completely broken. There can be special orbits satisfying the constraint and belonging to strata with a non-trivial little group, but these orbits do not represent generic
solutions of the constraint. The region defined by the constraint is invariant under $A_4$ transformations and the theory can be regarded as a non-linear realisation of the discrete symmetry $A_4$.\hfill
%%%%%%%%%%%%%%%%%%%%%%%%%%%%%%%%%%%%%%%%%%%%%%%%%%%%%%%%%%%%%%%%%%%%%%%%%%%%%%%%%%%%%%%%%%%
%%%%%%%%%%%%%%%%%%%%%%%%%%%%%%%%%%%%%%%%%%%%%%%%%%%%%%%%%%%%%%%%%%%%%%%%%%%%%%%%%%%%%%%%%%%
\section{Discussion}
%%%%%%%%%%%%%%%%%%%%%%%%%%%%%%%%%%%%%%%%%%%%%%%%%%%%%%%%%%%%%%%%%%%%%%%%%%%%%%%%%%%%%%%%%%%
We have proposed a new bottom-up approach to the problem of lepton masses and mixing angles based on supersymmetric modular invariant theories.
By recalling general properties of such theories we have seen how they naturally involve the finite modular groups $\Gamma_N$. When chiral multiplets transform, up to an overall factor, in
unitary representations of $\Gamma_N$, the Yukawa couplings have to be modular forms of level $N$.
Modular invariance plays the role of flavour symmetry.
Flavons field might not be needed and the flavour symmetry can be entirely broken by the VEV of the modulus. The superpotential of the theory is very restricted and modular invariance constrains both neutrino masses and mixing angles.
As long as supersymmetry is exact all higher-dimensional operators in the superpotential are completely determined by modular invariance. Possible sources of corrections are supersymmetry-breaking contributions 
and modifications of the Kahler potential. When the weights of all matter supermultiplets vanish, the theory collapse to a supersymmeric $\Gamma_N$-invariant model,
of the type that has been extensively studied in the past.
We have provided several complete examples based on $\Gamma_3$, by explicitly constructing the ring of modular forms of level 3. 
In the most economical example, discussed in section 3.1.2, neutrino mass ratios, lepton mixing angles, Dirac and Majorana phases 
do not depend on any Lagrangian parameter, but only on the modulus VEV. In the difficult challenge  of reproducing five known observables in terms of a single complex parameter,
this model scores relatively well. Only $\sin^2\theta_{13}$, predicted to be about $0.045$, is out of the allowed experimental range,
while providing a reasonable zeroth-order approximation. The model predicts inverted neutrino mass ordering.
The example presented in section 3.1.3 predicts normal neutrino mass ordering and is almost equally well-performing.
The last example, discussed in section 3.1.4, is minimal in the sense that there are no additional fields breaking the flavour symmetry other than the modulus.
Lepton mass ratios and the CP-violating phase are in agreement with the observations,
but the mixing angles are only qualitatively reproduced. Solar and atmospheric angles are large and the reactor angle is small, but they are out of range by many standard deviations.
Nevertheless the example provides an additional proof that this type of constructions can be carried out and that they are worth exploring. 

When discussing modular forms of level 3 we came across a new feature, allowing to extend the notion of non-linearly realised symmetry to the discrete case.
While in the continuous case non-linear realizations can be defined by imposing invariant constraints on the field space, in the discrete case
we can also consider covariant constraints of the type $\chi_i(\varphi)=0$, $\chi_i$ denoting non-invariant singlets of the group,
a possibility that is precluded to (connected, semisimple) Lie groups. These conditions can be used to
define non-linear realizations of a discrete group and we have explicitly analyzed how they can arise for the groups $S_3$ and $A_4$. 
Non-linear realizations of a discrete symmetry do not define low-energy effective
theories. They should rather be view as consistent truncations of some ultraviolet completion. In general covariant constraints leaves no non-trivial
residual symmetry. We identify the cases in which a non-trivial residual symmetry group survives, by discussing the properties of 
the so-called orbit space, the space spanned by the invariants of the group.  
The discussion given here about  non-linearly realized discrete symmetries is very provisional
and has been motivated by a puzzle about the dimension of the linear spaces of modular forms.
Nevertheless we hope that in the future it may turn into a useful tool when looking for new possibilities in model building.
%%%%%%%%%%%%%%%%%%%%%%%%%%%%%%%%%%%%%%%%%%%%%%%%%%%%%%%%%%%%%%%%%%%%%%%%%%%%%%%%%%%%%%%%%%%
\section*{Aknowledgements}
%%%%%%%%%%%%%%%%%%%%%%%%%%%%%%%%%%%%%%%%%%%%%%%%%%%%%%%%%%%%%%%%%%%%%%%%%%%%%%%%%%%%%%%%%%%
I am grateful to Stefano Forte, Aharon Levy and Giovanni Ridolfi  for giving me the chance to honour the memory of Guido Altarelli with this contribution.
I thank Franc Cameron for useful correspondence about modular forms and Gianguido Dall'Agata and Roberto Volpato for comments on the manuscript. This  work  was  supported  in  part  by  the  MIUR-PRIN  project  2015P5SBHT 003  ``Search for the Fundamental Laws and Constituents'' and  by  the
European  Union network FP10  ITN ELUSIVES  and INVISIBLES-PLUS (H2020-  MSCA-
ITN- 2015-674896 and H2020- MSCA- RISE- 2015- 690575). 
%%%%%%%%%%%%%%%%%%%%%%%%%%%%%%%%%%%%%%%%%%%%%%%%%%%%%%%%%%%%%%%%%%%%%%%%%%%%%%%%%%%%%%%%%%%
\section*{Appendix A}
%%%%%%%%%%%%%%%%%%%%%%%%%%%%%%%%%%%%%%%%%%%%%%%%%%%%%%%%%%%%%%%%%%%%%%%%%%%%%%%%%%%%%%%%%%%
We list here general formulas for the dimensions $d_{2k}(\Gamma(N))$ of the linear space ${\cal M}_{2k}(\Gamma(N))$ of modular forms of level $N$ and weight $2k$ \cite{gunning}. The dimensions read
\be
d_{2k}(\Gamma)=
\left\{
\begin{array}{ll}
\left[k/6\right]& k=1~({\tt mod}~6)\\[0.2 cm]
\left[k/6\right]+1& k\ne 1~({\tt mod}~6)
\end{array}
\right.
\label{dGamma}
\ee
where $[x]$ is the integer part of $x$, and
\be
\left\{
\begin{array}{l}
d_{2k}(\Gamma(2))=k+1\\[0.2 cm]
d_{2k}(\Gamma(N))=\dd\frac{(2k-1)N+6}{24}N^2\prod_{p|N} (1-\frac{1}{p^2})~~~~~~~~~N>2
\end{array}
\right.
\ee
where the product is over the prime divisors $p$ of $N$. 
%%%%%%%%%%%%%%%%%%%%%%%%%%%%%%%%%%%%%%%%%%%%%%%%%%%%%%%%%%%%%%%%%%%%%%%%%%%%%%%%%%%%%%%%%%%
\section*{Appendix B}
%%%%%%%%%%%%%%%%%%%%%%%%%%%%%%%%%%%%%%%%%%%%%%%%%%%%%%%%%%%%%%%%%%%%%%%%%%%%%%%%%%%%%%%%%%%
We show that modular forms $f_i(\tau)$ of weight $2k$ and level $N\ge 2$ transform under $\Gamma_N$ as
\be
f_i(\gamma \tau)=(c' \tau+d')^{2k} \rho(\gamma)_{ij}f_j(\tau)
\label{ttt}
\ee
where
\be
\gamma=\left(\begin{array}{cc}a'&b'\\c'&d'\end{array}\right)
\ee
is representative of an element in $\Gamma_N$ and $\rho(\gamma)$ is a unitary representation of $\Gamma_N$.
It is sufficient to show
that (\ref{ttt}) holds for $\gamma$ equal to $S$ and $T$, which generate the entire modular group $\overline{\Gamma}$ and are not elements of $\overline{\Gamma}(N)$ for $N\ge 2$.
When $\gamma$ is equal to $S$ and $T$  eq. (\ref{ttt}) becomes
\be
\left\{
\begin{array}{l}
(S\tau)^{2k} f_i(S \tau)=\rho(S)_{ij}f_j(\tau)\\[0.2 cm]
f_i(T \tau)=\rho(T)_{ij}f_j(\tau)
\end{array}
\right.~~~.
\label{repST}
\ee
We start by observing that the holomorphic functions $F_{Ti}(\tau)=f_i(T \tau)$ and $F_{Si}(\tau)=(S\tau)^{2k}f_i(S \tau)$ are modular forms of weight $2k$ and level $N$.
Let $h$
\be
h=\left(\begin{array}{cc}a&b\\c&d\end{array}\right)
\ee
  be a generic element of $\Gamma(N)$.  Making use of the fact that $\Gamma(N)$ are normal subgroups of $\Gamma$, we have:
  \be
  F_{Ti}(h\tau)=f_i(ThT^{-1} T\tau)=(c \tau+d)^{2k} F_{Ti}(\tau)
  \nn
  \ee
  and
  \be
  \begin{array}{l}
  F_{Si}(h\tau)=(Sh\tau)^{2k} f_i(ShS^{-1}S\tau)= (c\tau+d)^{2k}(S\tau)^{2k} f_i(S\tau)=(c\tau+d)^{2k} F_{Si}(\tau)
  \end{array}~~~.
  \nn
  \ee
Thus $f_i(T \tau)$ and $(S\tau)^{2k}f_i(S \tau)$ are in ${\cal M}_{2k}(\Gamma(N))$ and can be written as linear combinations of $f_j(\tau)$, like in eq. (\ref{repST}). The fact that the coefficients $\rho_{ij}(S)$ and $\rho_{ij}(T)$ define representations of $\Gamma_N$
follows from the algebraic relations satisfied by $S$ and $T$. For instance, the generators $S$ and $T$ of $\Gamma_3$ satisfy the relations:
\be
S^2=T^3=(ST)^3=\mathds{1}~~~.
\ee
It follows that:
\be
\begin{array}{l}
\rho_{ik}(S)\rho_{kj}(S)=\rho_{ik}(T)\rho_{km}(T)\rho_{mj}(T)=\delta_{ij}\\
\rho_{ik}(S)\rho_{km}(T)\rho_{mp}(S)\rho_{pq}(T)\rho_{qr}(S)\rho_{rj}(T)=\delta_{ij}~~~,
\end{array}
\ee
and $\rho_{ij}(S)$ and $\rho_{ij}(T)$ are linear representations of the generators of $\Gamma_3$. The presentations of several groups $\Gamma_N$ in terms of the elements $S$ and $T$ can be found in ref. \cite{deAdelhartToorop:2011re}. 
Finally, the unitarity of the representation stems from the finite dimensionality of $\Gamma_N$.
%%%%%%%%%%%%%%%%%%%%%%%%%%%%%%%%%%%%%%%%%%%%%%%%%%%%%%%%%%%%%%%%%%%%%%%%%%%%%%%%%%%%%%%%%%%
\section*{Appendix C}
%%%%%%%%%%%%%%%%%%%%%%%%%%%%%%%%%%%%%%%%%%%%%%%%%%%%%%%%%%%%%%%%%%%%%%%%%%%%%%%%%%%%%%%%%%%
$A_4$ is the group of even permutations of four objects. It is also the symmetry group of a regular tetrahedron. It has 12 elements and two generators, $S$ and $T$, satisfying
\be
S^2=T^3=(ST)^3=1~~~.
\label{presA4}
\ee
The 12 elements fall into four conjugacy classes $C_1=\{e\}$, $C_2=\{T,ST,TS,STS\}$, $C_3=\{T^2,ST^2,T^2S,TST\}$, $C_4=\{S,T^2ST,TST^2\}$. The group has four irreducible representations: an invariant singlet $1$, two non-invariant singlets $1',1''$ and a triplet $3$.
The elements $S$ and $T$ in the irreducible representations of $A_4$ are shown in table \ref{irred}. The one-dimensional representations are determined non-ambiguously by the
conditions (\ref{presA4}), while the three-dimensional representation is determined up to a unitary transformation, representing a change of basis. The last two columns
of table 9 display $S$ and $T$ in two convenient basis, the real one and the complex one, respectively.
\begin{table}[h] 
\centering
\begin{tabular}{|c|c|c|c|c||c|}
\hline
&$1$& $1'$& $1''$ & $3$ ({\tt Real})& $3$ ({\tt Complex})\\
\hline
$S$&$1$ &$1$ &$1$ &
$\left(
\begin{array}{ccc}
1&0&0\\
0&-1&0\\
0&0&-1
\end{array}
\right)$ &
$\dd\frac{1}{3}\left(
\begin{array}{ccc}
-1&2&2\\
2&-1&2\\
2&2&-1
\end{array}
\right)$\rule[-4ex]{0pt}{10ex} \\
\hline
$T$&$1$ &$\omega$ &$\omega^2$ &
$\left(
\begin{array}{ccc}
0&1&0\\
0&0&1\\
1&0&0
\end{array}
\right)$
&
$\left(
\begin{array}{ccc}
1&0&0\\
0&\omega&0\\
0&0&\omega^2
\end{array}
\right)$\rule[-4ex]{0pt}{10ex} \\
\hline
\end{tabular}
\caption{Elements $S$ and $T$ in the irreducible representations of $A_4$, $\omega=-1/2+i \sqrt{3}/2$ denoting a cubic root of unity. For the triplet representation, two possible choice of basis are displayed.}
\label{irred}
\end{table}
Given two triplets $\varphi=(\varphi_1,\varphi_2,\varphi_3)$ and $\psi=(\psi_1,\psi_2,\psi_3)$, their product decomposes in the sum $1+1'+1''+3_S+3_A$,
where $3_{S(A)}$ denotes the symmetric(antisymmetric) combination. We show the result in table \ref{decomp}, both in the real and in the complex basis.
\begin{table}[h] 
\centering
\begin{tabular}{|c|c|c|}
\hline
&{\tt Real basis}&{\tt Complex basis}\\
\hline
$(\varphi\psi)_1$&$\varphi_1\psi_1+\varphi_2\psi_2+\varphi_3\psi_3$&$\varphi_1\psi_1+\varphi_2\psi_3+\varphi_3\psi_2$\rule[-2ex]{0pt}{5ex}\\
\hline
$(\varphi\psi)_{1'}$&$\varphi_1\psi_1+\omega^2\varphi_2\psi_2+\omega\varphi_3\psi_3$ &$\varphi_3\psi_3+\varphi_1\psi_2+\varphi_2\psi_1$\rule[-2ex]{0pt}{5ex}\\
\hline
$(\varphi\psi)_{1''}$&$\varphi_1\psi_1+\omega\varphi_2\psi_2+\omega^2\varphi_3\psi_3$ &$\varphi_2\psi_2+\varphi_3\psi_1+\varphi_1\psi_3$\rule[-2ex]{0pt}{5ex}\\
\hline
$(\varphi\psi)_{3_S}$&$\left(
\begin{array}{c}\varphi_2\psi_3+\varphi_3\psi_2\\ \varphi_3\psi_1+\varphi_1\psi_3\\ \varphi_1\psi_2+\varphi_2\psi_1\end{array}\right)$ &$\dd\frac{1}{\sqrt{3}}\left(
\begin{array}{c}2\varphi_1\psi_1-\varphi_2\psi_3-\varphi_3\psi_2\\2\varphi_3\psi_3-\varphi_1\psi_2-\varphi_2\psi_1 \\ 2\varphi_2\psi_2-\varphi_3\psi_1-\varphi_1\psi_3 \end{array}\right)$\rule[-4ex]{0pt}{10ex}\\
\hline
$(\varphi\psi)_{3_A}$&$\left(
\begin{array}{c}\varphi_2\psi_3-\varphi_3\psi_2\\ \varphi_3\psi_1-\varphi_1\psi_3\\ \varphi_1\psi_2-\varphi_2\psi_1\end{array}\right)$ &$\left(
\begin{array}{c}\varphi_2\psi_3-\varphi_3\psi_2\\ \varphi_1\psi_2-\varphi_2\psi_1 \\ \varphi_3\psi_1-\varphi_1\psi_3 \end{array}\right)$\rule[-4ex]{0pt}{10ex}\\
\hline
\end{tabular}
\caption{Decomposition of the product of two triplets $\varphi=(\varphi_1,\varphi_2,\varphi_3)$ and $\psi=(\psi_1,\psi_2,\psi_3)$ in irreducible representations, both for the real and the complex basis.}
\label{decomp}
\end{table}
%%%%%%%%%%%%%%%%%%%%%%%%%%%%%%%%%%%%%%%%%%%%%%%%%%%%%%%%%%%%%%%%%%%%%%%%%%%%%%%%%%%%%%%%%%%
\section*{Appendix D}
%%%%%%%%%%%%%%%%%%%%%%%%%%%%%%%%%%%%%%%%%%%%%%%%%%%%%%%%%%%%%%%%%%%%%%%%%%%%%%%%%%%%%%%%%%%
We explicitly construct a basis of weight 2 modular forms for $\Gamma(3)$.
The dimension of the space $M_2(\Gamma(3))$ of modular forms of weight 2 for $\Gamma(3)$ has dimension 3 and we look for three linearly elements.
We start by observing that if $f(\tau)$ transforms as
\be
f(\tau)\to e^{i\alpha}~(c\tau+d)^k~f(\tau)~~~,
\nn
\ee
then
\be
\frac{d}{d\tau}\log f(\tau)\to (c\tau+d)^2 \frac{d}{d\tau}\log f(\tau)+ k~c(c\tau+d)~~~.
\nn
\ee
The inhomogeneous term can be removed if we combine several $f_i(\tau)$ with weights $k_i$
\be
\frac{d}{d\tau}\sum_i\log f_i(\tau)\to (c\tau+d)^2 \frac{d}{d\tau}\sum_i\log f_i(\tau)+ \left(\sum_i k_i\right) c(c\tau+d)~~~,
\nn
\ee
provided the sum of the weights vanishes. Consider the Dedekind eta-function $\eta(\tau)$, defined in the upper complex plane:
\be
\eta(\tau)=q^{1/24}\prod_{n=1}^\infty \left(1-q^n \right)~~~~~~~~~~~~~q\equiv e^{i 2 \pi\tau}~~~.
\nn
\ee
It satisfies
\be
\eta(-1/\tau)=\sqrt{-i \tau}~\eta(\tau)~~~,~~~~~~~~~~\eta(\tau+1)=e^{i \pi/12}~ \eta(\tau)~~~.
\nn
\ee
So $\eta(\tau)^{24}$ is a modular form of weight 12 under the full modular group, actually a cusp form. 
We further observe that the set composed by $\eta(3\tau)$, $\eta(\tau/3)$, $\eta((\tau+1)/3)$ and $\eta((\tau+2)/3)$
is closed under the modular group. Under $T$ we have
\bea
\eta(3\tau)&\to& e^{\dd i \pi/4}~ \eta(3\tau)\nn\\
\eta\left(\frac{\tau}{3}\right)&\to&\eta\left(\frac{\tau+1}{3}\right)\nn\\
\eta\left(\frac{\tau+1}{3}\right)&\to&\eta\left(\frac{\tau+2}{3}\right)\nn\\
\eta\left(\frac{\tau+2}{3}\right)&\to&e^{\dd i \pi/12}~\eta\left(\frac{\tau}{3}\right)~~~.\nn
\eea
Under $S$ we have
\bea
\eta(3\tau)&\to&\sqrt{1/3}~\sqrt{-i \tau}~\eta\left(\frac{\tau}{3}\right) \nn\\
\eta\left(\frac{\tau}{3}\right)&\to&\sqrt{3}~\sqrt{-i \tau}~\eta(3\tau)\nn\\
\eta\left(\frac{\tau+1}{3}\right)&\to&e^{\dd- i \pi/12}~\sqrt{-i \tau}~\eta\left(\frac{\tau+2}{3}\right)\nn\\
\eta\left(\frac{\tau+2}{3}\right)&\to&e^{\dd i \pi/12}~\sqrt{-i \tau}~\eta\left(\frac{\tau+1}{3}\right)~~~.\nn
\eea
A candidate weight 2 form is
\be
Y(\alpha,\beta,\gamma,\delta\vert\tau)=\frac{d}{d\tau}\left[\alpha\log\eta\left(\frac{\tau}{3}\right)+\beta\log\eta\left(\frac{\tau+1}{3}\right)+\gamma\log\eta\left(\frac{\tau+2}{3}\right)+\delta\log\eta(3\tau)
\right]~~~.
\nn
\ee
with $\alpha+\beta+\gamma+\delta=0$ to eliminate the inhomogeneous term. Under $S$ and $T$ we have
\be
Y(\alpha,\beta,\gamma,\delta\vert\tau)\xrightarrow{S} \tau^2~ Y(\delta,\gamma,\beta,\alpha\vert\tau)~~~,~~~~~~~~~~Y(\alpha,\beta,\gamma,\delta\vert\tau)\xrightarrow{T} Y(\gamma,\alpha,\beta,\delta\vert\tau)~~~.
\nn
\ee
We search for three independent forms $Y_i(\tau)$ transforming in the three-dimensional representation of $A_4$  (in a vector notation where $Y^T=(Y_1,Y_2,Y_3)$)
\be
Y(-1/\tau)=\tau^2~\rho(S) Y(\tau)~~~,~~~~~~~~~~Y(\tau+1)=\rho(T) Y(\tau)~~~,
\nn
\ee
with unitary matrices
$\rho(S)$ and $\rho(T)$
\vskip 0.1cm
\be
\rho(S)=\frac{1}{3}
\left(
\begin{array}{ccc}
-1&2&2\\
2&-1&2\\
2&2&-1
\end{array}
\right)~~~,~~~~~~~~~~
\rho(T)=
\left(
\begin{array}{ccc}
1&0&0\\
0&\omega&0\\
0&0&\omega^2
\end{array}
\right)~~~,~~~~~~~\omega=-\frac{1}{2}+\frac{\sqrt{3}}{2}i~~~.
\nn
\ee
\vskip 0.1cm
\noindent
The  transformation $T$ requires:
\be
Y_1(\tau)=c_1~Y(1,1,1,-3\vert\tau)~~~,~~~~~~~Y_2(\tau)=c_2~Y(1,\omega^2,\omega,0\vert\tau)~~~,~~~~~~~Y_3(\tau)=c_3~Y(1,\omega,\omega^2,0\vert\tau)~~~,
\nn
\ee
while the transformation $S$ fixes the coefficients $c_i$ up to an overall factor:
\be
c_1=3~c~~~,~~~~~~~~~~c_2=-6~c~~~,~~~~~~~~~~c_3=-6~c~~~.
\nn
\ee
For convenience we choose $c=i/2\pi$ and explicitly we have
\bea
Y_1(\tau)&=&\frac{i}{2 \pi}
   \left[
   \frac{\eta'\left(\frac{\tau }{3}\right)}{\eta \left(\frac{\tau}{3}\right)}
   +\frac{\eta'\left(\frac{\tau+1}{3}\right)}{\eta \left(\frac{\tau+1}{3}\right)}
   +\frac{\eta'\left(\frac{\tau +2}{3}\right)}{\eta\left(\frac{\tau +2}{3}\right)}
   -\frac{27 \eta'(3 \tau )}{\eta (3 \tau)}
   \right]\nn\\
Y_2(\tau)&=&\frac{-i}{\pi}
   \left[
   \frac{\eta'\left(\frac{\tau }{3}\right)}{\eta \left(\frac{\tau}{3}\right)}
+\omega^2~\frac{\eta'\left(\frac{\tau+1}{3}\right)}{\eta \left(\frac{\tau+1}{3}\right)}
   +\omega~\frac{\eta'\left(\frac{\tau +2}{3}\right)}{\eta\left(\frac{\tau +2}{3}\right)}
   \right]\\
Y_2(\tau)&=&\frac{-i}{\pi}
   \left[
   \frac{\eta'\left(\frac{\tau }{3}\right)}{\eta \left(\frac{\tau}{3}\right)}
+\omega~\frac{\eta'\left(\frac{\tau+1}{3}\right)}{\eta \left(\frac{\tau+1}{3}\right)}
   +\omega^2~\frac{\eta'\left(\frac{\tau +2}{3}\right)}{\eta\left(\frac{\tau +2}{3}\right)}
   \right]~~~.\nn
\eea
The $q$-expansion of $Y_i(\tau)$ reads:
\bea
Y_1(\tau)&=&1+12q+36q^2+12q^3+...\nn\\
Y_2(\tau)&=&-6q^{1/3}(1+7q+8q^2+...)\nn\\
Y_3(\tau)&=&-18q^{2/3}(1+2q+5q^2+...)~~~.\nn
\eea
It agrees, up to an overall factor, with the $q$ expansion derived in ref. \cite{franc}, where the functions $Y_i(\tau)$ are expressed in terms of hypergeometric series. 
From the $q-$expansion we see that the modular forms $Y_i(\tau)$ satisfy the constraint:
\be
(YY)_{1''}\equiv Y_2^2+2 Y_1 Y_3=0~~~.
\label{c3}
\ee
Notice that this constraint is left invariant by $A_4$ transformations, since the combination $(YY)_{1''}$ is invariant up to a phase factor.
This constraint has a direct consequence on the number of non-vanishing multilinear combinations of $Y_i$. If we do not require eq. (\ref{c3}), there are $(k+1)(k+2)/2$ independent $k$-linear combinations of the type
\be
Y_{i_1}Y_{i_2} \cdot\cdot\cdot Y_{i_k}~~~~~~~(i_1,i_2,i_k=1,2,3)~~~.
\label{multil}
\ee
For instance, for $k=2$, we have the 6 combinations $Y_1 Y_1$, $Y_2 Y_1$, $Y_3 Y_1$, $Y_2 Y_2$, $Y_3 Y_2$, $Y_3 Y_3$. These combinations can be arranged into irreducible representations
of $A_4$. For $k=2$ the six independent $Y_iY_j$ decompose as $3+1+1'+1''$ under $A_4$:
\bea
Y^{(4)}_3&=&(Y_1^2-Y_2 Y_3,Y_3^2-Y_1 Y_2,Y_2^2-Y_1 Y_3)\nn\\
Y^{(4)}_1&=&Y_1^2+2 Y_2 Y_3\nn\\
Y^{(4)}_{1'}&=&Y_3^2+2 Y_1 Y_2\\
Y^{(4)}_{1''}&=&Y_2^2+2 Y_1 Y_3~~~,\nn
\eea
where, in the left-hand side, the lower index denotes the representation and the upper index stands for $2 k$. 
For $k=3$, we have the 10 independent combinations $Y_i Y_j Y_l$.
It is not difficult to see that they should decompose as $3+3+3+1$ under $A_4$. They are
\bea
Y^{(6)}_1&=&Y_1^3+Y_2^3+Y_3^3-3 Y_1 Y_2 Y_3\nn\\
Y^{(6)}_{3,1}&=&(Y_1^3+2 Y_1Y_2Y_3,Y_1^2Y_2+2Y_2^2Y_3,Y_1^2Y_3+2Y_3^2Y_2)\nn\\
Y^{(6)}_{3,2}&=&(Y_3^3+2 Y_1Y_2Y_3,Y_3^2Y_1+2Y_1^2Y_2,Y_3^2Y_2+2Y_2^2Y_1)\nn\\
Y^{(6)}_{3,3}&=&(Y_2^3+2 Y_1Y_2Y_3,Y_2^2Y_3+2Y_3^2Y_1,Y_2^2Y_1+2Y_1^2Y_3)\nn
\eea
As a consequence of the constraint in eq. (\ref{c3}) we see that $Y^{(4)}_{1''}=0$ and $Y^{(6)}_{3,3}=0$. In general, $k(k-1)/2$ among the
$Y_{i_1}Y_{i_2} \cdot\cdot\cdot Y_{i_k}$ are equal to zero and we are left with $2k+1$ nonvanishing combinations. These should still decompose into $A_4$ representations, since the constraint (\ref{c3}) is $A_4$-invariant. 
This explains why the linear space of modular forms of level 3 and weight $2k$ has dimension $2k+1$, as it should, and not $(k+1)(k+2)/2$ as naively derived by counting the homogeneous polynomials of degree $k$ in $Y_i(\tau)$.
Thus a modular invariant supersymmetric theory using as building blocks modular forms of level 3, as the one we have explicitly illustrated as a candidate model for lepton masses,
can be regarded as a sort of non-linear realisation of the discrete symmetry $A_4$.
%%%%%%%%%%%%%%%%%%%%%%%%%%%%%%%%%%%%%%%%%%%%%%%%%%%%%%%%%%%%%%%
 
 %%%%%%%%%%%%%%%%%%%%%%%%%%%%%%%%%%%%%%%%%%%%%%%%%%%%%%%%%%%%%%%%

\end{document}